\documentclass[a4paper,11pt]{article}
\pdfoutput=1 

\usepackage{jcappub} 

\usepackage[T1]{fontenc} 

\makeatletter
\def\@fpheader{\relax}
\makeatother

\title{\boldmath Quadratic, Higgs and hilltop potentials in the Palatini gravity}

\author[]{Nilay Bostan}

\affiliation{Department of Physics and Astronomy, University of Iowa, \\52242, Iowa City, IA, USA }

\emailAdd{nilay-bostan@uiowa.edu}

\abstract{In this work, we study inflation with the non-minimally coupled quadratic, Standard Model (SM) Higgs and hilltop potentials through $\xi \phi^2R$ term in the Palatini gravity. We first analyze observational parameters of Palatini quadratic potential as functions of $\xi$ for high-$N$ scenario and low-$N$ scenario. In addition to this, taking into account inflaton $\phi$ has a non-zero vacuum expectation value $v$ afterwards inflation, we display observational parameters of well-known symmetry-breaking potentials type of Higgs potential and its generalizations which are hilltop potentials in the Palatini formalism for high-$N$ scenario and low-$N$ scenario. We calculate inflationary parameters of Palatini Higgs potential as functions of $v$ for different $\xi$ values where inflaton values both $\phi>v$ and $\phi<v$ during inflation as well as we show that observational parameters of Palatini Higgs potential in the induced gravity limit for high-$N$ scenario. On the other hand, we illustrate different from the Higgs potential the effect of $\xi$ on hilltop potentials which can agree with the observations for inflaton value solely $\phi<v$ and $\xi$, $v\ll1$ for both two scenarios, which we mentioned above. For each considered potentials, we also display $n_s-r$ values fit the current data given by the Keck Array/BICEP2 and Planck collaborations.}


\begin{document}
\maketitle
\flushbottom

\section{Introduction}
Inflation \cite{Guth:1980zm,Linde:1981mu,Albrecht:1982wi,Linde:1983gd} is the most outstanding scenario for the primordial universe. It is considered that inflation has become a solution isotropy and homogeneity problem as well as it explains spatial flatness of the Universe to a high degree successfully. This inflationary era can also produce and extend the small inhomogeneities which have appeared in the large scale structures and the anisotropy in the cosmic microwave background radiation temperature (CMBR). The most recent measurements of the CMBR \cite{Aghanim:2018eyx,Akrami:2018odb} made by the Planck satellite give some parameters that are related with the inflationary perturbations two have already become even more precisely in recent years which are the amplitude of the curvature perturbation, $\Delta_\mathcal{R}^2\approx  2.4\times10^{-9}$ and the corresponding spectral index, $n_s= 0.9625\pm 0.0048$. The other parameter which is namely as the running of the spectral index, $\alpha=0.002\pm 0.010 $. Even though the current constraints on the $\alpha$ are not sufficient to test the inflationary models, they are considered to be enhanced much with observations of the 21 cm line \cite{Kohri:2013mxa,Basse:2014qqa,Munoz:2016owz} approximately at the level of $\alpha=\mathcal{O} (10^{-3})$. In addition to this, the recent data from the Keck Array/BICEP2 and Planck collaborations \cite{Ade:2018gkx} constraints strongly the tensor-to-scalar ratio $r<0.06$, which gives successful explanation to the amplitude of primordial gravitational waves and the scale of inflation. Some ongoing CMB B-mode polarization experiments \cite{Wu:2016hul,Matsumura:2013aja,Ade:2018sbj} pushed the limit decrease to $r\lesssim0.001$ or targeting to detect up this limit. Each parameters above are constrained at the pivot scale $k_* = 0.002$ Mpc$^{-1}$.  

The observational parameters, in particular the spectral index $n_s$ and the tensor-to-scalar ratio $r$ have been calculated for various inflationary potentials \cite{Martin:2013tda}. However, the most minimal realization scenario of inflation is the Standard Model Higgs boson behaves as the inflaton field with minimal coupling ($\xi=0$). On the other hand, a renormalizable scalar field theory in curved space-time needs to the non-minimal coupling $\xi\phi^2R$ between the inflaton and the Ricci scalar \cite{Callan:1970ze,Freedman:1974ze,Buchbinder:1992rb}. Furthermore, even if the non-minimal coupling $\xi$ equals to zero at the classical level, it will be created by quantum corrections \cite{Callan:1970ze} and in particular, non-minimal coupling to gravity is necessary to sufficiently flatten the Higgs potential at large field values, so that it is agreement with observations. In this paper aims to extend the previous studies of inflaton is coupled non-minimally to gravity, presenting how the value of the non-minimal coupling parameter $\xi$ affects the observational parameters for the inflationary potentials in the Palatini formalism, in case of the quadratic potential and symmetry-breaking type inflation potentials where inflaton has a non-zero vacuum expectation value $v$ afterwards inflation. Non-zero $v$ after inflation is that such potentials can be related with symmetry-breaking in the very early universe. Examples of such models for symmetry-breaking which we investigate in this work the well-known Higgs inflation \cite{Salopek:1988qh,Bezrukov:2007ep} models which are based on Standard Model of particle physics and especially, the Higgs field of the SM behaves as the inflaton field, a scenario proposed by ref. \cite{Bezrukov:2007ep}. We also discuss hilltop potentials which are simple generalization of Higgs potential. 

Furthermore, in this paper, we use dynamics of the Palatini gravity to be able to calculate inflationary parameters. Although the Metric and Palatini formalisms are equivalent in the theory of General Relativity, if matter fields are coupled non-minimally to gravity, these two formalisms correspond to two different theories of gravity such refs. investigated \cite{Bauer:2008zj,York:1972sj,Tenkanen:2017jih,Rasanen:2017ivk,Racioppi:2017spw,Tamanini:2010uq}. In particular, inflationary models with non-minimal couplings to gravity can not be explained only form of potential, gravitational degrees of freedom requires to define \cite{Bauer:2008zj}. In the Palatini formalism different from Metric one, both the metric $g_{\mu \nu}$ and the connection $\Gamma$ are independent variables. Even though the two formalisms have the same equations of motion and as a result they correspond to the equivalent physical theories, the presence of the non-minimal coupling between gravity and matter, physical equivalence is disappear for these two formalisms, in particular the $\xi$-attractor models which are known as attractor behavior occuring to the Starobinsky model for larger $\xi$ values in Metric formulation is lost in the Palatini approach \cite{Kallosh:2013tua} and $r$ can be taken much smaller values compared to the Metric formulation for larger $\xi$ values \cite{Racioppi:2017spw,Barrie:2016rnv,Kannike:2015kda,Artymowski:2016dlz}. Another different case is between Metric and Palatini formalism, the inflaton stays sub-Planckian regime to supply a natural inflationary era in the Palatini formalism \cite{Bauer:2008zj}. 

In literature, inflationary potentials in Palatini gravity are taken into account some papers \cite{Tenkanen:2017jih,Bauer:2008zj,Rasanen:2017ivk,Jinno:2019und,Rubio:2019ypq,Enckell:2018kkc}. In ref. \cite{Tenkanen:2017jih} discussed quadratic potential in Palatini gravity taking $N_*=50$ and $N_*=60$, they found that strength of non-minimal coupling, $\xi=\mathcal{O} (10^{-3})$ to agree with the current data just for $N_*=60$. In addition to this, Higgs inflation in Palatini formulation has been studied refs. \cite{Bauer:2008zj,Rasanen:2017ivk,Jinno:2019und,Rubio:2019ypq,Enckell:2018kkc}. According to these papers, predictions of $r$ is very tiny for $\xi\gtrsim1$ values and so $r$ is highly suppressed further  well-known attractor behaviour in the Metric formulation for large $\xi$ values in Starobinsky model is vanished for Palatini approach. The paper is organized as follows, we first describe inflation with a non-minimal coupling and how inflationary parameters calculate (section \ref{non}) in Palatini formulation. Next, we analyze Palatini quadratic potential in the large-field limit (section \ref{quadraticpot}). We then calculate inflationary predictions in detail for two different symmetry-breaking inflation type potentials, known as the Higgs potential (section \ref{higgs}) for inflaton value for both $\phi>v$ and $\phi<v$ as well as different from the Higgs potential, we illustrate hilltop potentials can compatible with the current measurements for cases of $\phi<v$ and $\xi$, $v\ll1$ (section \ref{hilltop}). Furthermore, (in section \ref{higgs}) we calculate inflationary parameters in the induced gravity limit for Palatini Higgs inflation. Finally, we discuss our results and summary of them (section \ref{conc}).  
\section{Palatini inflation with a non-minimal coupling} \label{non}
We describe non-minimally coupled scalar field $\phi$  with a canonical kinetic term and a potential $V_J(\phi)$ inflation action in the Jordan frame 
\begin{equation}\label{nonminimal_action}
S_J = \int d^4x \sqrt{-g}\left(\frac{1}{2}F(\phi) g^{\mu\nu}R_{\mu\nu}(\Gamma) - \frac{1}{2} g^{\mu\nu}\partial_{\mu}\phi\partial_{\nu}\phi - V_J(\phi) \right). 
\end{equation}
Here, the subscript $J$ indicates that the action is defined in a Jordan frame. $R_{\mu\nu}$ is the Ricci tensor and it is defined by that form
\begin{equation}\label{Riccitensor}
R_{\mu\nu}=\partial_{\sigma}\Gamma_{\mu \nu}^{\sigma}-\partial_{\mu}\Gamma_{\sigma \nu}^{\sigma}+\Gamma_{\mu \nu}^{\rho}\Gamma_{\sigma \rho}^{\sigma}-\Gamma_{\sigma \nu}^{\rho}\Gamma^{\sigma}_{\mu \rho}.
\end{equation}
In the metric formulation the connection is taken as a function of metric tensor, that is, Levi-Civita connection  ${\bar{\varGamma}={\bar{\varGamma}}(g^{\mu\nu})}$
\begin{equation} \label{vargammametric}
\bar{\varGamma}_{\mu\nu}^{\lambda}=\frac{1}{2}g^{\lambda \rho} (\partial_{\mu}g_{\nu \rho}+\partial_{\nu}g_{\rho \mu}-\partial_{\rho}g_{\mu\nu}).
\end{equation}
On the other hand, in the Palatini formalism both $g_{\mu \nu}$ and $\varGamma$ are independent variables, and the only assumption that the connection is torsion-free, i.e. $\varGamma_{\mu\nu}^{\lambda}=\varGamma_{\nu\mu }^{\lambda}$. If solving equations of motion, one can be obtained as follows \cite{Bauer:2008zj}
\begin{equation}\label{vargammapalatini}
\Gamma^{\lambda}_{\mu\nu} = \overline{\Gamma}^{\lambda}_{\mu\nu}
+ \delta^{\lambda}_{\mu} \partial_{\nu} \omega(\phi) +
\delta^{\lambda}_{\nu} \partial_{\mu} \omega(\phi)\nonumber\\ - g_{\mu \nu} \partial^{\lambda} \omega(\phi),
\end{equation}
where 
\begin{eqnarray}
\label{omega}
\omega\left(\phi\right)=\ln\sqrt{F(\phi)},
\end{eqnarray}
in the Palatini formulation. In this work, in order to calculate inflationary parameters of symmetry-breaking type inflation potentials, we choose $F(\phi)$ includes of a constant $m^2$ term and a non-minimal coupling $\xi \phi^2 R$ which is necessary for renormalizable scalar field theory in curved space-time \cite{Callan:1970ze,Freedman:1974ze,Buchbinder:1992rb} as we mentioned above. We are using units that the reduced Planck scale $m_P=1/\sqrt{8\pi
G}\approx2.4\times10^{18}\text{ GeV}$ is set equal to unity, thus we consider $F(\phi)\to1$ or $\phi\to0$ after inflation. In that case, by taking into consideration $m^2=1-\xi v^2$, we obtain $F(\phi)=m^2+\xi \phi^2=1+\xi(\phi^2-v^2) $ \cite{Bostan:2018evz}. What is more, we take into account Palatini quadratic potential in the large-field limit, so to be able to compute observational parameters, we take $F(\phi)=1+\xi \phi^2$. 

\subsection{Calculating the inflationary parameters}
The difference between Metric and Palatini formulations are more easily figured out in the Einstein frame by applying a Weyl rescaling $g_{E, \mu \nu}=g_{\mu \nu}/F(\phi)$ and thus  Einstein frame action displaying in that form

\begin{eqnarray}\label{einsteinframe}
S_E = \int d^4x \sqrt{-g_{E}}\left(\frac{1}{2}g_E^{\mu\nu}R_{E, \mu \nu}(\Gamma)-\frac{1}{2Z(\phi)}\, g_E^{\mu\nu} \partial_{\mu}\phi\partial_{\nu}\phi - \frac{V_E(\phi)}{F(\phi)^2} \right),
\end{eqnarray}
where
\begin{equation} \label{Zphi}
Z^{-1}(\phi)=\frac{1}{F(\phi)},
\end{equation}
in the Palatini formulation. If we make a field redefinition
\begin{equation}\label{redefine}
\mathrm{d}\chi=\frac{\mathrm{d}\phi}{\sqrt{Z(\phi)}}\,,
\end{equation}
we obtain the action for a minimally coupled scalar field $\chi$ with a canonical kinetic term. Using eq. \eqref{redefine}, Einstein frame action in terms of $\chi$ is given that form
\begin{eqnarray}\label{einsteinframe}
S_E = \int d^4x \sqrt{-g_{E}}\left(\frac{1}{2}g_E^{\mu\nu}R_{E}(\Gamma)-\frac{1}{2}\, g_E^{\mu\nu} \partial_{\mu}\chi\partial_{\nu}\chi - V_E(\chi) \right).
\end{eqnarray}

For $F(\phi)=1+\xi(\phi^2-v^2)$, eq. \eqref{Zphi} can be defined with different limit cases:
\begin{description}
	\item[1.] Electroweak regime \\ 
	If $|\xi(\phi^2-v^2)|\ll1$,  $\phi\approx\chi$ and
	$V_J(\phi)\approx V_E(\chi)$. Thus, the inflationary
	predictions are approximately the same as for minimal coupling case. 
	\item[2.] Induced gravity limit \cite{Zee:1978wi} \\
	In this limit ($\xi v^2=1$, $F(\phi)=\xi\phi^2$),  $Z(\phi)=\xi\phi^2$ and using eq. \eqref{redefine}, we obtain
	\begin{equation}\label{induced}
	\phi=v\exp\left(\chi\sqrt{\xi}\right)\,,
	\end{equation}
	here we took $\chi(v)=0$.
	\item[3.] Large-field limit \\
	If $\phi^2\gg v^2$ during inflation, we have
	\begin{equation} \label{strong}
	\phi\simeq\frac{1}{\sqrt{\xi}}\sinh \left(\chi\sqrt{\xi}\right),
	\end{equation}
	in the Palatini formulation. Using eq. \eqref{strong}, inflationary potential can be taken into account in terms of canonical scalar field $\chi$, therefore slow-roll parameters are written for Palatini formulation in the large-field limit according to $\chi$.
\end{description}

On the condition that Einstein frame potential is written in terms of the canonical scalar field $\chi$, inflationary parameters can be found using the slow-roll parameters
\cite{Lyth:2009zz}
\begin{equation}\label{slowroll1}
\epsilon =\frac{1}{2}\left( \frac{V_{\chi} }{V}\right) ^{2}\,, \quad
\eta = \frac{V_{\chi\chi} }{V}  \,, \quad
\xi ^{2} = \frac{V_{\chi} V_{\chi \chi\chi} }{V^{2}}\,,
\end{equation}
where $\chi$'s in the subscript represent derivatives. Inflationary parameters can be defined in the slow-roll approximation by
\begin{eqnarray}\label{nsralpha1}
n_s = 1 - 6 \epsilon + 2 \eta \,,\quad
r = 16 \epsilon, \quad
\alpha=\frac{\mathrm{d}n_s}{\mathrm{d}\ln k} = 16 \epsilon \eta - 24 \epsilon^2 - 2 \xi^2\,.
\end{eqnarray}
In the slow-roll approximation, the number of e-folds is obtained by
\begin{equation} \label{efold1}
N_*=\int^{\chi_*}_{\chi_e}\frac{V\rm{d}\chi}{V_{\chi}}\,, \end{equation}
where the subscript ``$_*$'' indicates quantities when the scale
corresponding to $k_*$ exited the horizon, and $\chi_e$ is the inflaton
value at the end of inflation, which we obtain by $\epsilon(\chi_e) =
1$. 

The amplitude of the curvature perturbation in terms of canonical scalar field $\chi$ is written the form
\begin{equation} \label{perturb1}
\Delta_\mathcal{R}=\frac{1}{2\sqrt{3}\pi}\frac{V^{3/2}}{|V_{\chi}|}.
\end{equation}

Furthermore, we redefine slow-roll parameters in terms of scalar field $\phi$ for numerical calculations because for all general $\xi$ and $v$ values, it is not possible to compute the inflationary potential in terms of the $\chi$. Using with eqs. \eqref{redefine} and \eqref{slowroll1} together, slow-roll parameters can be found in terms of $\phi$ \cite{Linde:2011nh}
\begin{equation}\label{slowroll2}  
\epsilon=Z\epsilon_{\phi}\,,\quad
\eta=Z\eta_{\phi}+{\rm sgn}(V')Z'\sqrt{\frac{\epsilon_{\phi}}{2}}\,,\quad
\xi^2=Z\left(Z\xi^2_{\phi}+3{\rm sgn}(V')Z'\eta_{\phi}\sqrt{\frac{\epsilon_{\phi}}{2}}+Z''\epsilon_{\phi}\right)\,.
\end{equation}
where we described 
\begin{equation}
\epsilon_{\phi} =\frac{1}{2}\left( \frac{V^{\prime} }{V}\right) ^{2}\,, \quad
\eta_{\phi} = \frac{V^{\prime \prime} }{V}  \,, \quad
\xi ^{2} _{\phi}= \frac{V^{\prime} V^{\prime \prime\prime} }{V^{2}}\,.
\end{equation}
In addition to this, eqs. \eqref{efold1} and \eqref{perturb1} can be obtained in terms of $\phi$ in that form
\begin{eqnarray}\label{perturb2}
N_*&=&\rm{sgn}(V')\int^{\phi_*}_{\phi_e}\frac{\mathrm{d}\phi}{Z(\phi)\sqrt{2\epsilon_{\phi}}}\,,\\
\label{efold2} \Delta_\mathcal{R}&=&\frac{1}{2\sqrt{3}\pi}\frac{V^{3/2}}{\sqrt{Z}|V^{\prime}|}\,.
\end{eqnarray}

\begin{figure}[h!]
	\centering
	\includegraphics[angle=0, width=12cm]{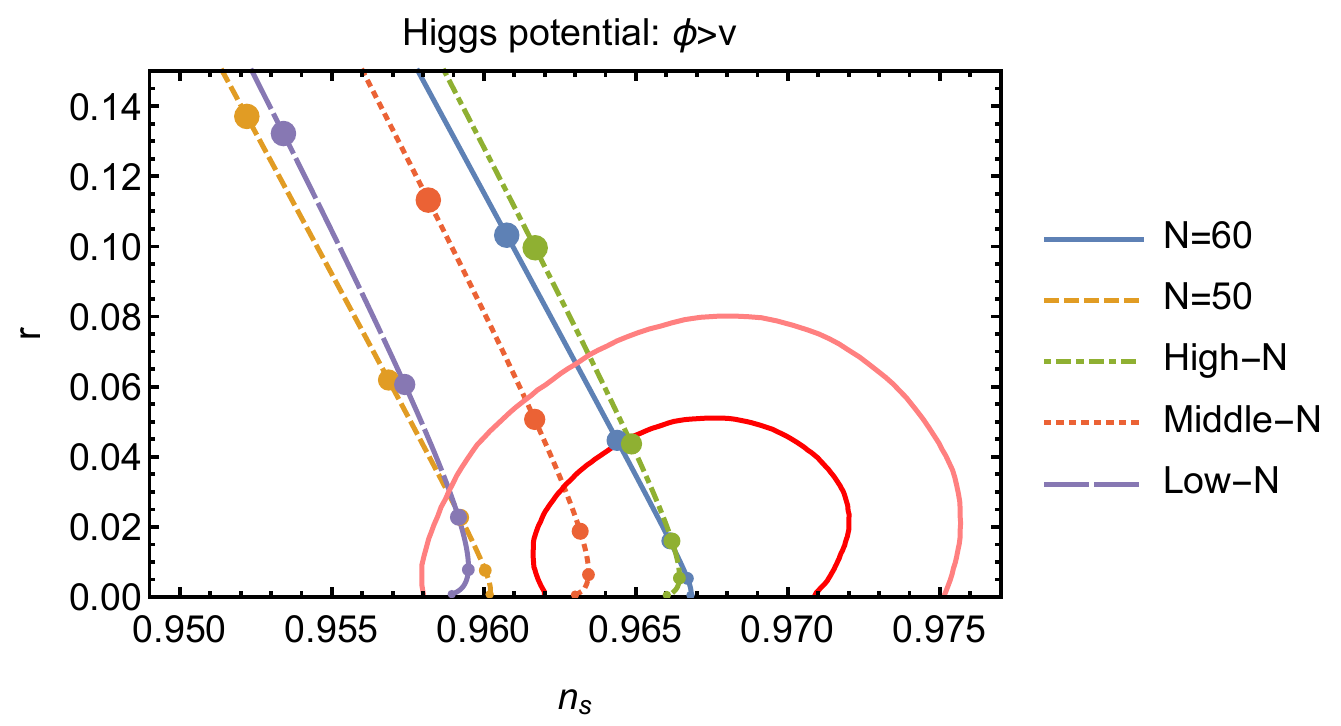}
	\caption{The figure illustrates that $n_s$-$r$ predictions for different $\xi$ values and $v=0.01$ for various reheating cases as described in the text for Higgs potential in the Palatini formalism. The points on each curve represent to $\xi=10^{-2.5},\,10^{-2},\,10^{-1.5},\,10^{-1},$ and $\,1$, top to bottom. The pink
		(red) contour corresponds to the 95\% (68\%) CL contours
		based on data taken by the Keck Array/BICEP2 and Planck collaborations \cite{Ade:2018gkx}.}
	\label{fig1}
\end{figure}

To compute the values of inflationary parameters, we should obtain
a value of $N_*$ numerically. Supposing that a standard thermal history after inflation, $N_*$ is given as follows \cite{Liddle:2003as}
\begin{equation} \label{efolds}
N_*\approx64.7+\frac12\ln\frac{\rho_*}{m^4_P}-\frac{1}{3(1+\omega_r)}\ln\frac{\rho_e}{m^4_P}
+\left(\frac{1}{3(1+\omega_r)}-\frac14\right)\ln\frac{\rho_r}{m^4_P}\,.
\end{equation}
Here $\rho_{e}=(3/2)V(\phi_{e})$ is the
energy density at the end of inflation, $\rho_*\approx V(\phi_*)$ is the energy density when the scale corresponding to $k_*$ exited the horizon. $\rho_r$ is the energy density at the
end of reheating and $\omega_r$ is the equation of state parameter throughout
reheating, which we take its value to be constant. Inflationary parameters predictions change depending on the total number of e-folds. In literature, most of papers take between $N_*\approx 50-60$ to be constant calculating to the inflationary parameters in general. On the other hand, to be able to discriminate inflationary models from each other, their predictions should know accurately. Therefore, to indicate an acceptable range of $N_*$ depending upon reheating temperature, we take into account three different scenario to define $N_*$: 
\begin{description}
	\item[1.] High-$N$ scenario\\ 
	$\omega_{r}=1/3$, this case corresponds to
	assuming instant reheating. 
	\item[2.] Middle-$N$ scenario \\
	$\omega_{r}=0$
	and the temperature of reheating is taken $T_r=10^9$ GeV, computing $\rho_r$ using the
	SM value for the usual number of relativistic degrees of freedom values for $g_*=106.75$.
	\item[3.] Low-$N$ scenario \\
	$\omega_{r}=0$ same as middle-$N$ scenario but in this case, the reheat temperature $T_r=100$ GeV. 
\end{description}
The $n_s-r$ curve for different scenarios are displayed in figure \ref{fig1} for the Higgs
potential in the Palatini formulation (debated in section \ref{higgs}) together with the 
68\% and 95\% confidence level (CL) contours based on data taken by the Keck Array/BICEP2 and Planck collaborations \cite{Ade:2018gkx}. The figure illustrates that
for the Higgs potential in the Palatini formalism, the confidential $N_*$ values of 50 and 60 which are taken necessarily agreement with the range expected from a standard thermal history afterwards inflation. However, $N_*$ is
smaller (for example between roughly 45-55 providing that $v\sim0.01$) for the hilltop inflation models (described in section \ref{hilltop}) because inflation takes place at a lower energy scale in these models. 
\section{Quadratic potential} \label{quadraticpot}
\begin{figure}[h!]
	\centering
	\includegraphics[angle=0, width=14cm]{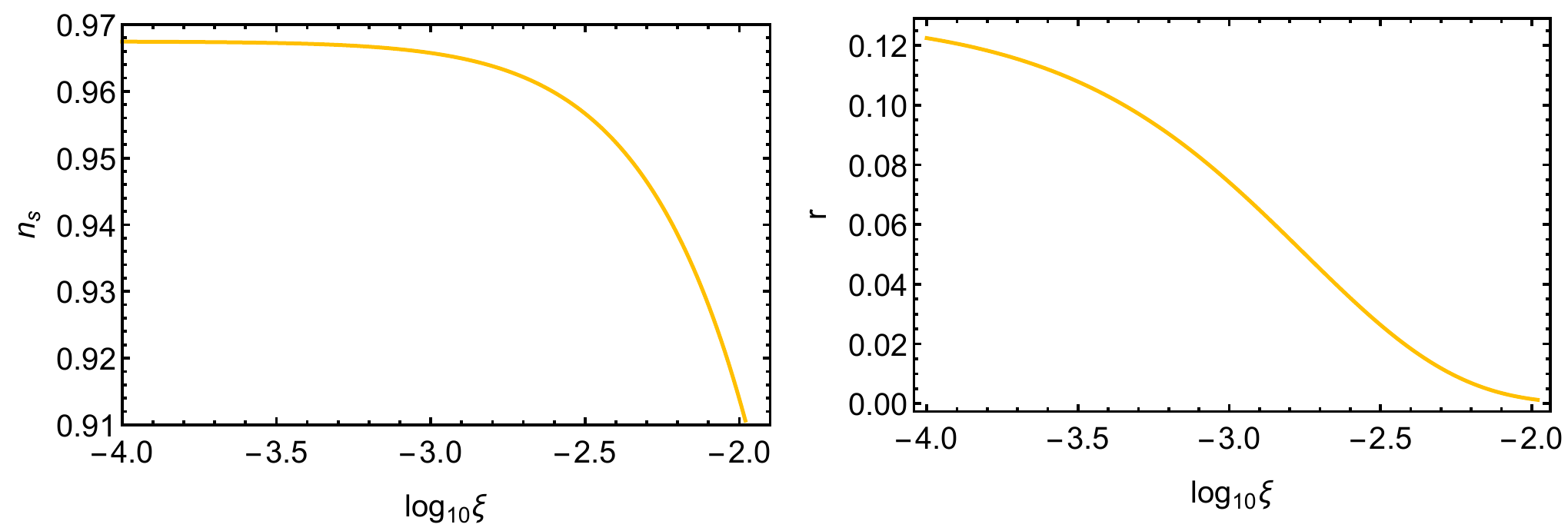}
	
	\
	
    \includegraphics[angle=0, width=9cm]{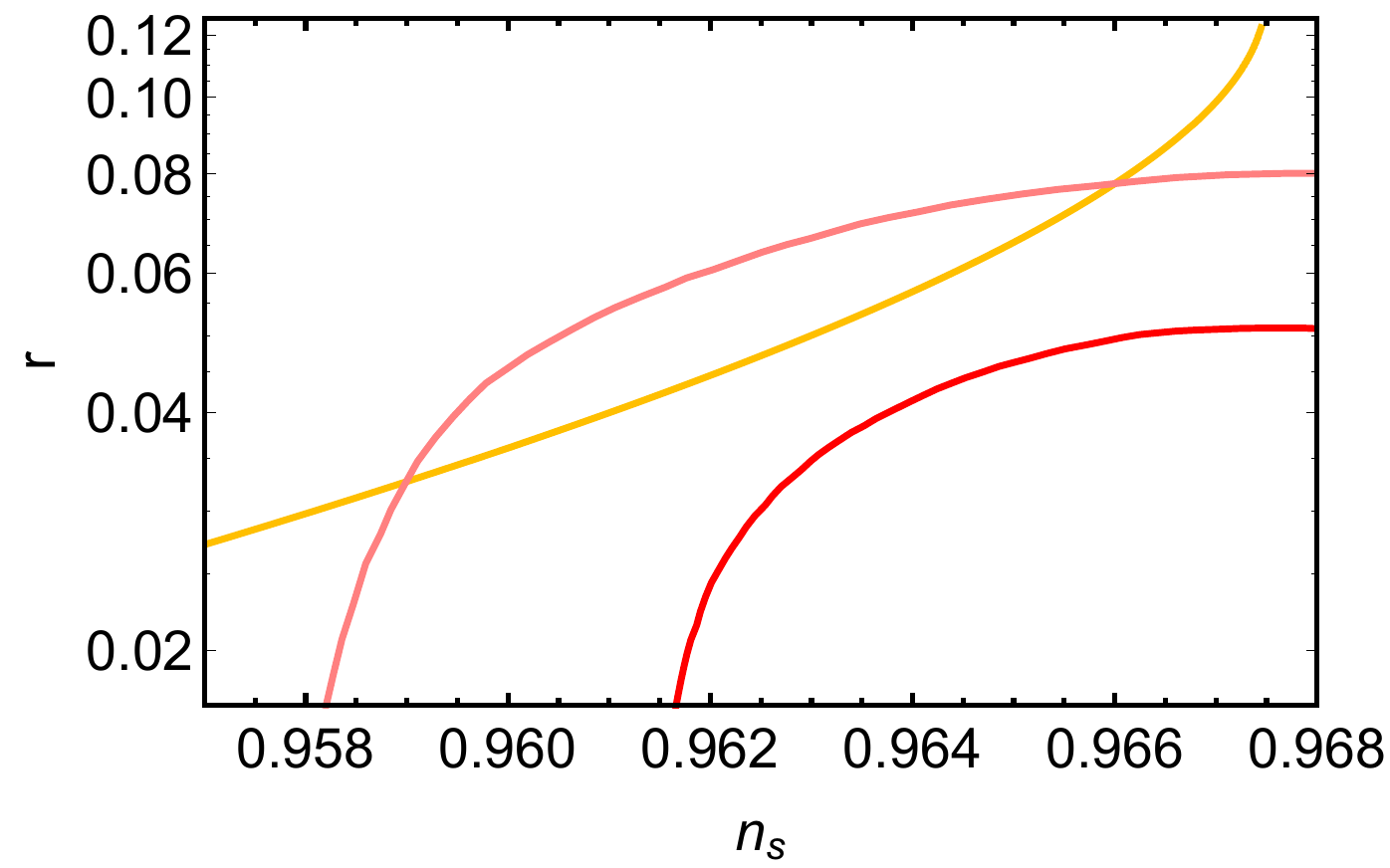}
	\caption{For quadratic potential in the Palatini formalism for high-$N$ scenario, top figures display that $n_s$, $r$ values as functions of $\xi$ and the bottom figure shows that $n_s-r$ predictions based on range of the top figures $\xi$ values. The pink (red) contour correspond to the 95\% (68\%) CL contour given by the Keck Array/BICEP2 and Planck collaborations \cite{Ade:2018gkx}.}
	\label{fig2}
\end{figure}

The quadratic inflation potential model in Jordan frame is given by in that form
\begin{equation}\label{quadraticpot1}
V_J(\phi)=\frac{1}{2}m^2\phi^2,
\end{equation}
here $m$ is a mass term and Einstein frame quadratic potential in the large-field limit (described in section \ref{non}) for Palatini approach in terms of $\chi$ using eq. \eqref{strong} can be obtained as follows
\begin{equation}\label{quadraticpot2}
V_E(\chi)\approx\frac{m^2}{2\xi}\frac{\sinh^2\left(\chi\sqrt{\xi}\right)}{\left(1+\sinh^2\left(\chi\sqrt{\xi}\right)\right)^2}.
\end{equation}
As it can be seen from eq. \eqref{quadraticpot2}, if expanding this potential around the minimum for large $\xi$ values, we can obtain flattening potential. In literature, ref. \cite{Tenkanen:2017jih} analyzed values of $n_s$, $r$ and $m$ for quadratic potential in Palatini gravity taking $N_*=50$ and $N_*=60$ to be constant. In this work, we analyze $n_s$, $r$, $\alpha$ and $m$ values as function of $\xi$ for Palatini quadratic potential with large-field limit numerically for high-$N$ scenario and low-$N$ scenario. According to our results from fig. \ref{fig2}, we find that if the non-minimal coupling parameter between the range $10^{-4}\lesssim\xi\lesssim 10^{-3}$ for high-$N$ scenario, values of $n_s$ can be inside observational region but in the case of larger $\xi$, $n_s$ values decrease and they remain outside the observational region as well as range between $10^{-4}\lesssim\xi\lesssim 10^{-2}$, we obtain $0.01\lesssim r\lesssim 0.12$.

\begin{figure}[h!]
	\centering
	\includegraphics[angle=0, width=7cm]{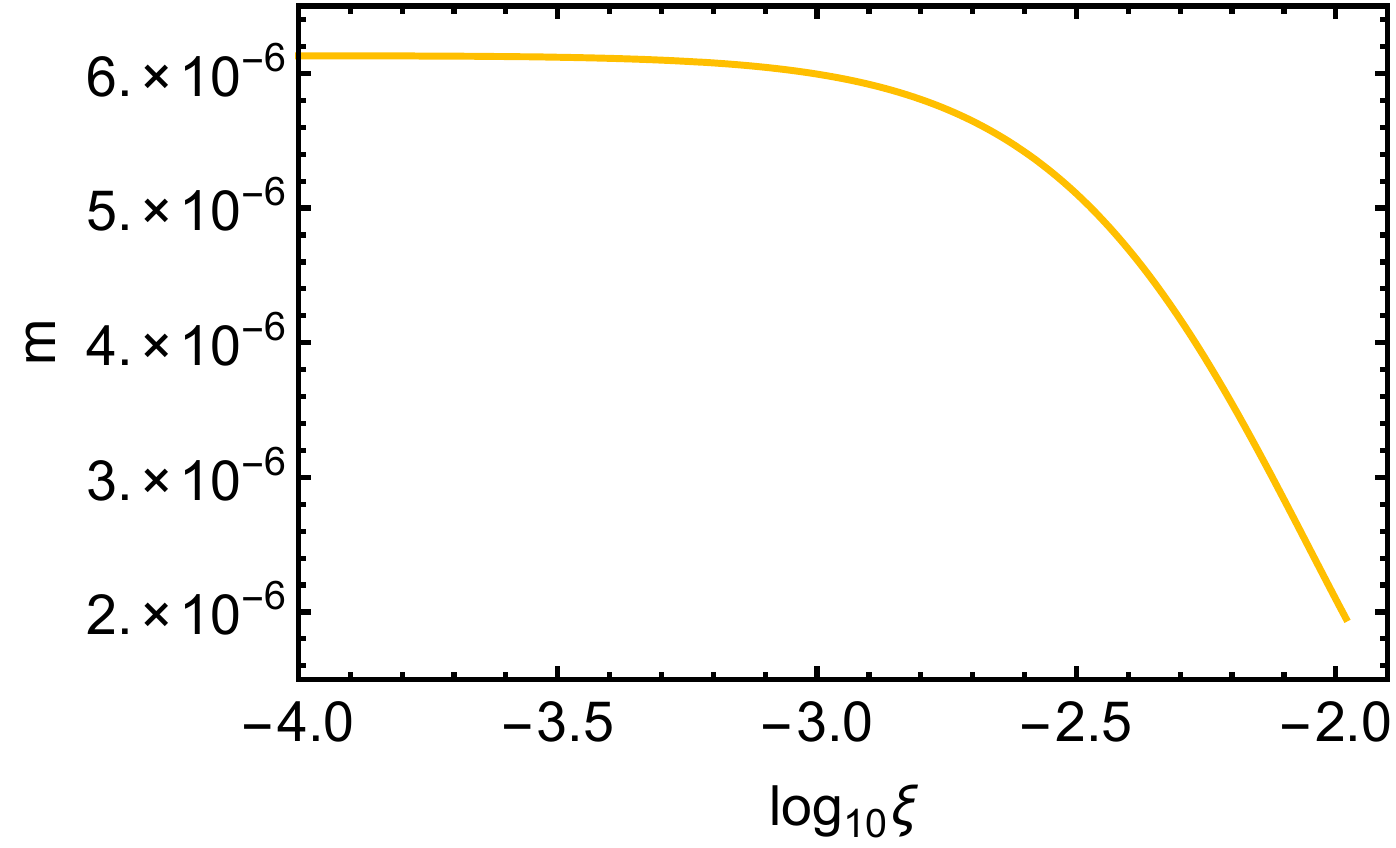}
	\includegraphics[angle=0, width=7cm]{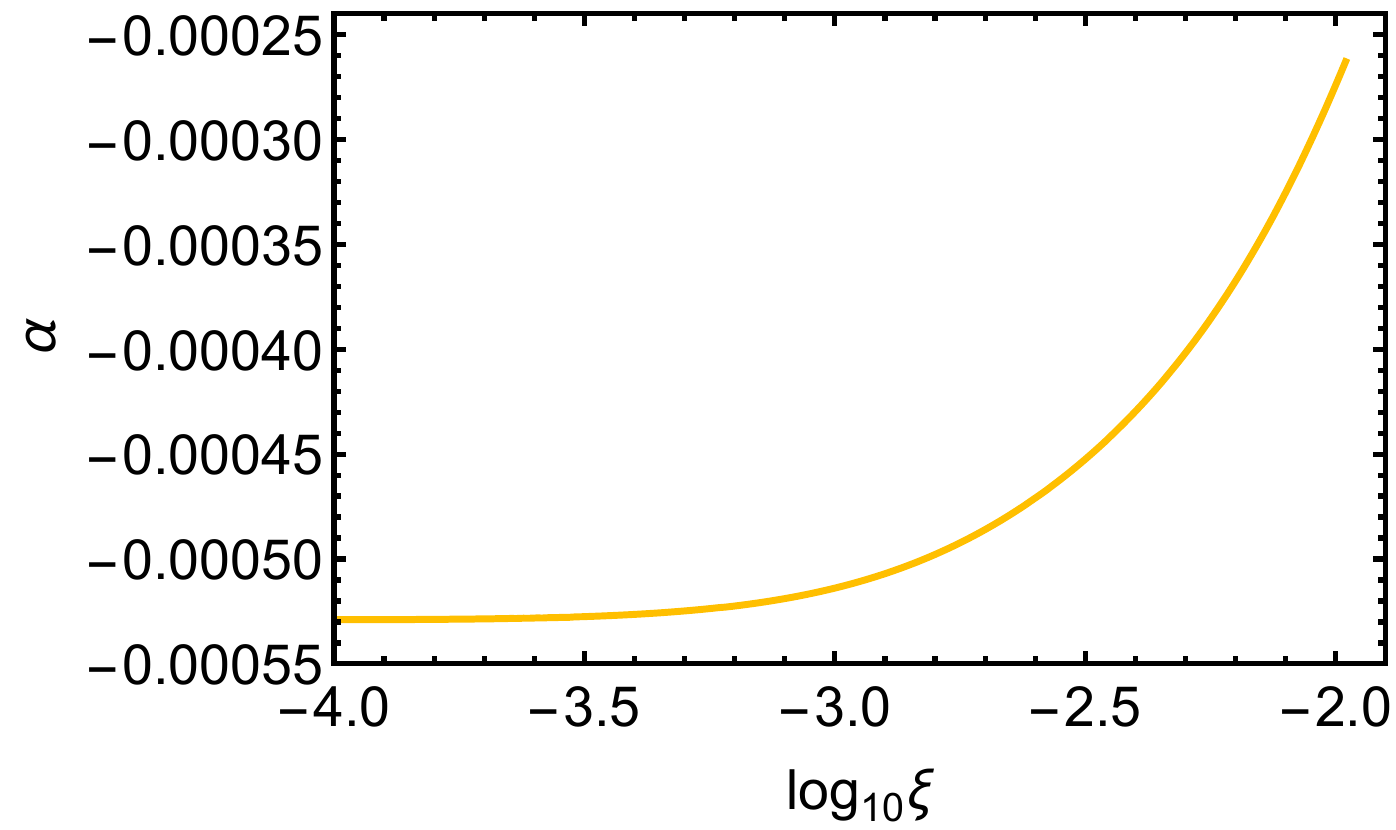}
	\caption{For quadratic potential in the Palatini formalism for high-$N$ scenario, the figures show that $m$ and $\alpha$ values as functions of $\xi$.}
	\label{fig3}
\end{figure}

\begin{figure}[h!]
	\centering
	\includegraphics[angle=0, width=14cm]{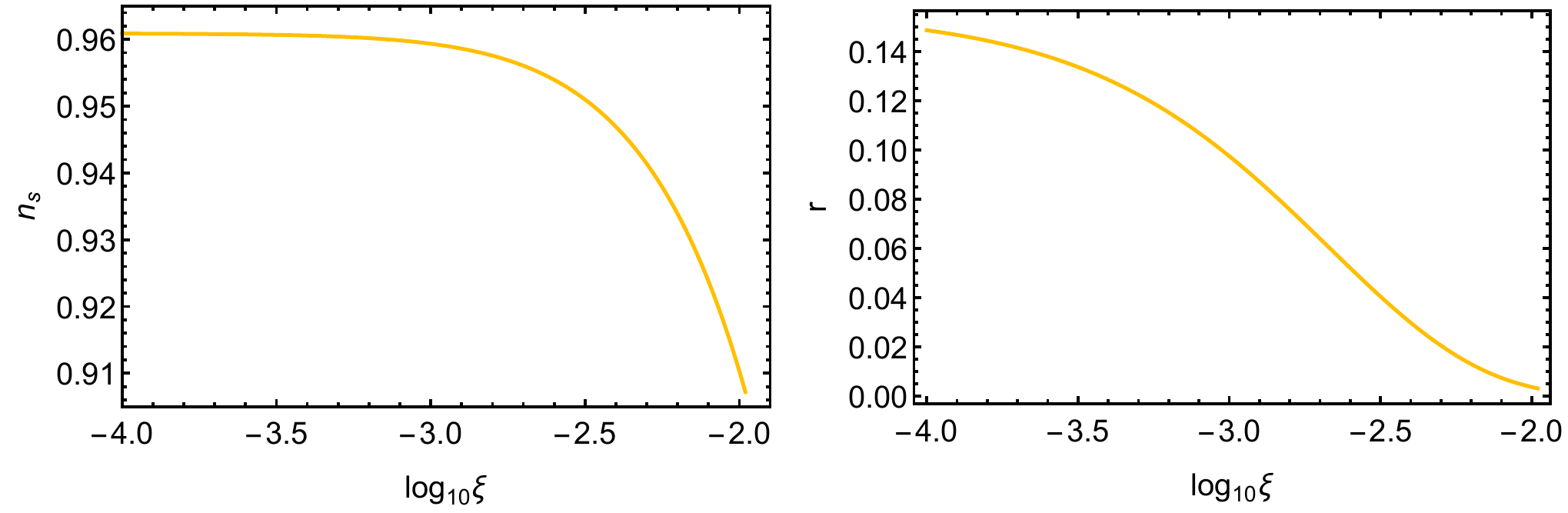}
	
	\
	
	\includegraphics[angle=0, width=9cm]{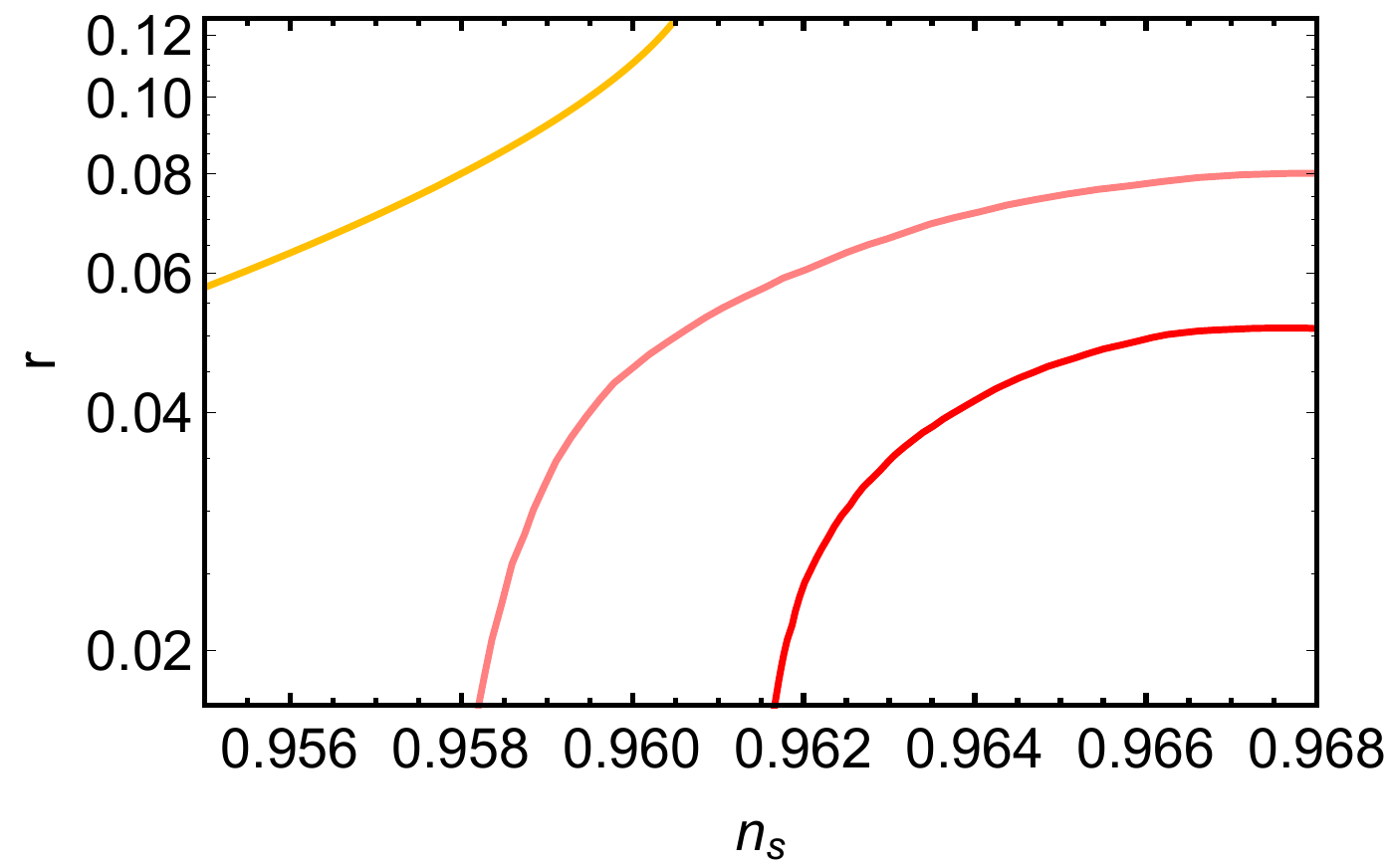}
	\caption{For quadratic potential in the Palatini formalism for low-$N$ scenario, top figures display that $n_s$, $r$ values as functions of $\xi$ and the bottom figure shows that $n_s-r$ predictions based on range of the top figures $\xi$ values. The pink (red) contour correspond to the 95\% (68\%) CL contour given by the Keck Array/BICEP2 and Planck collaborations \cite{Ade:2018gkx}.}
	\label{fig4}
\end{figure}

\begin{figure}[h!]
	\centering
	\includegraphics[angle=0, width=7cm]{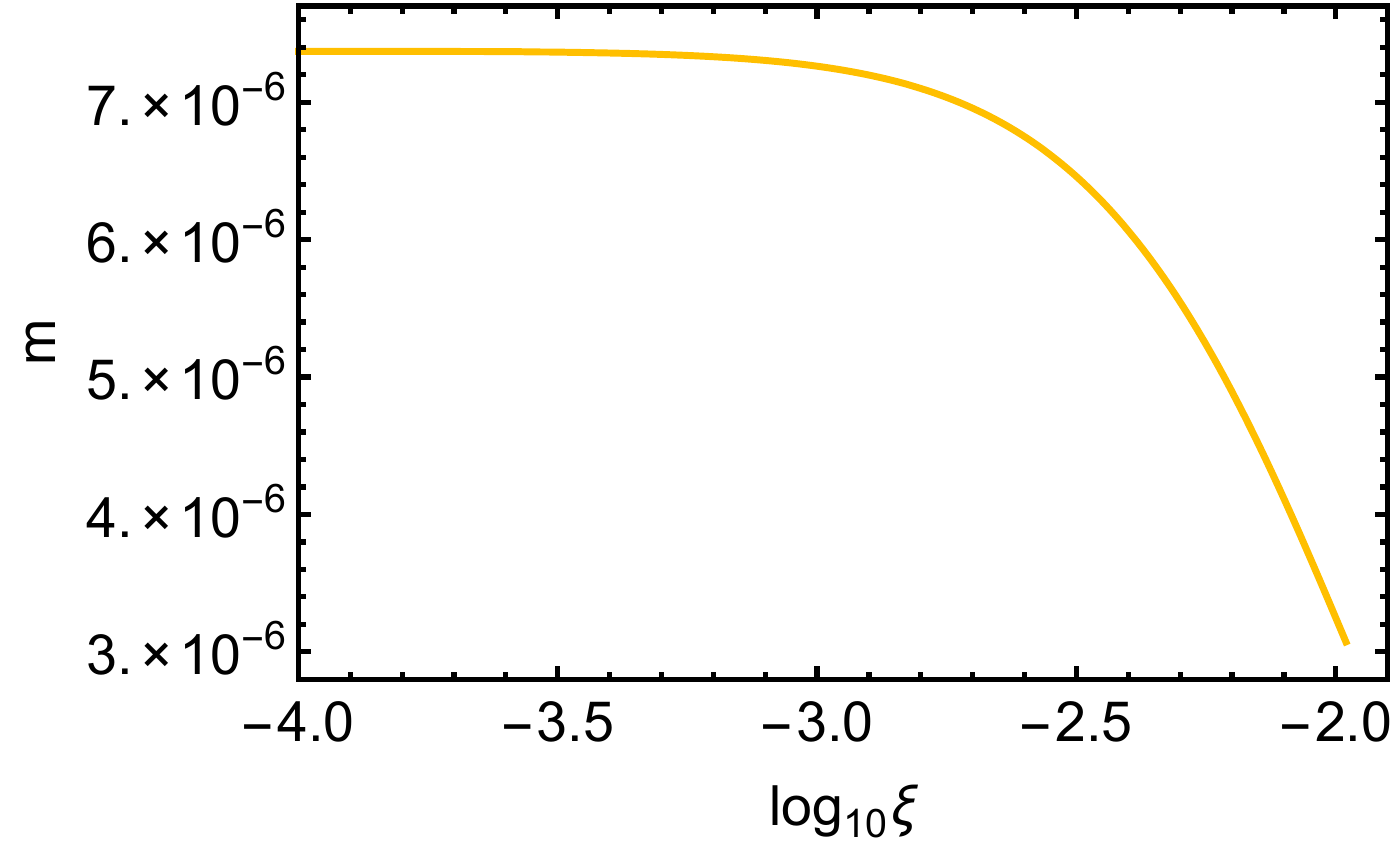}
	\includegraphics[angle=0, width=7cm]{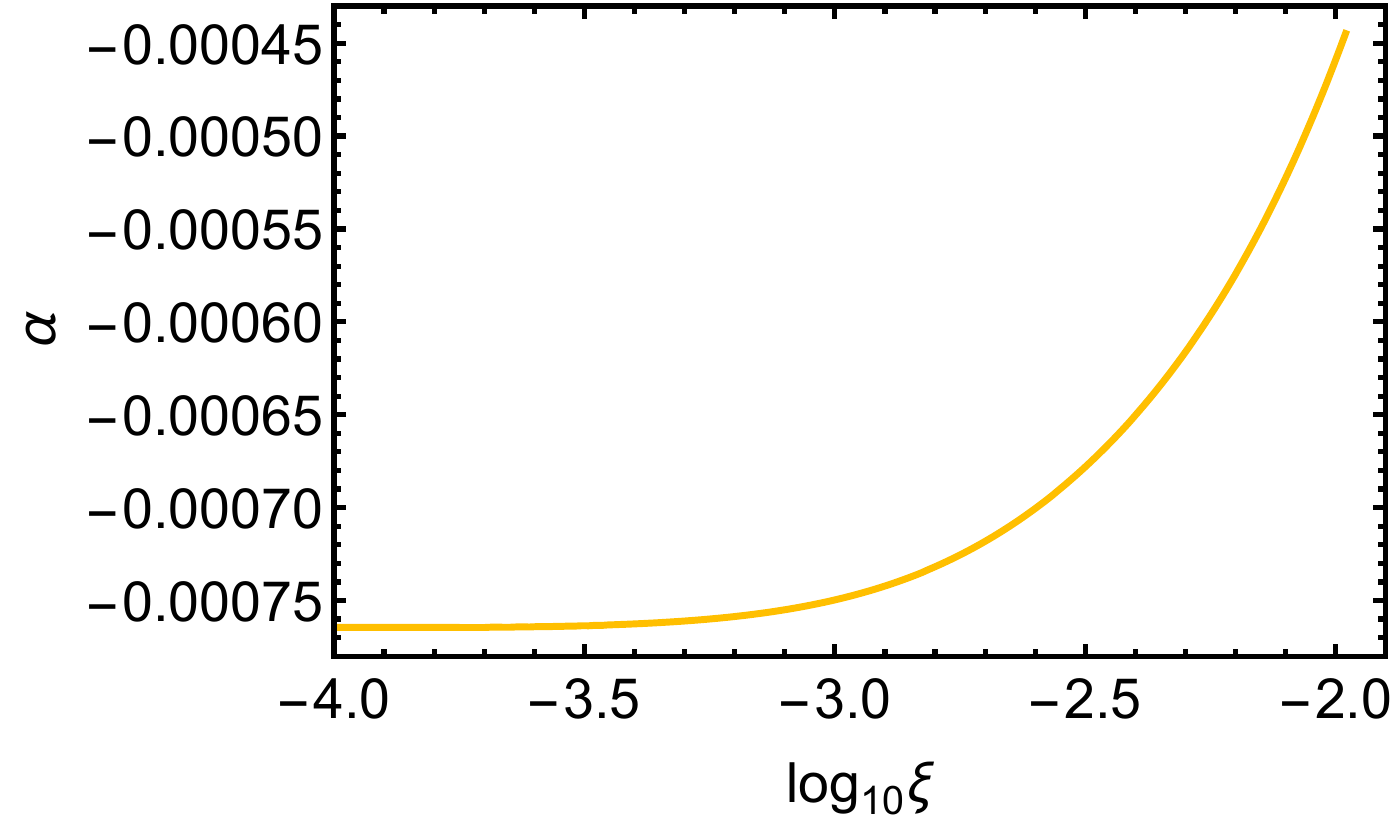}
	\caption{For quadratic potential in the Palatini formalism for low-$N$ scenario, the figures show that $m$ and $\alpha$ values as functions of $\xi$.}
	\label{fig5}
\end{figure}

On the other hand, for low-$N$ scenario (see fig. \ref{fig4}), values of $n_s$ are outside the observational region for any $\xi$ values and for between $10^{-4}\lesssim\xi\lesssim 10^{-2}$, we find $0.01\lesssim r\lesssim 0.14$ and also recent discussions about small $\xi$ values on same scenarios such refs. \cite{Alanne:2016mpa,Tenkanen:2016twd}. Furthermore, we show that $\alpha$ values are very small in Palatini quadratic potential for two different scenario to be able to observe near future experiments, as it can be seen from figs. \ref{fig3} and \ref{fig5} as well as in the observational region, value of $m$ for high-$N$ scenario approximately equals to $6\times 10^{-6}$ (see fig. \ref{fig3}) which may explain the reason why the SM Higgs ($m_H=125$ GeV) can not behave as an inflaton with quadratic potential. 

\

\

\clearpage
\section{Higgs potential} \label{higgs}
In this section, we take into account well-known symmetry-breaking type potential \cite{Goldstone:1961eq}
\begin{equation}\label{higgs1}
V_J(\phi)=A \left[1-\left(\frac{\phi}{v}\right)^2\right]^2\,,
\end{equation}
which is namely as Higgs potential. This potential for the minimal coupling case was investigated in recent such papers, see \cite{Vilenkin:1994pv,Linde:1994wt,Destri:2007pv,Martin:2013tda,Okada:2014lxa}. In this case, when inflation takes place around the minimum, the potential is approximately quadratic and thus the quadratic potential predictions in terms of $N_*$
\begin{equation}\label{quadratic}
n_s\approx1-\frac{2}{N_*}\,,\quad r\approx\frac{8}{N_*}\,,\quad
\alpha\approx-\frac{2}{N_*^2},
\end{equation}
can be obtained for inflation both $\phi>v$ and $\phi<v$. In this work, instead of minimal coupling case, we analyze Higgs inflation with non-minimal coupling in Palatini formulation both high-$N$ scenario and low-$N$ scenario.  Furthermore, using eq. \eqref{perturb2}, we obtain $N_*$ for non-minimally coupled Palatini Higgs inflation analytically in that form 
\begin{equation}\label{quad}
N_*=\frac{1}{8}\left( \phi_*^2-\phi_e^2\right)-\frac{v^2}{4}\ln\frac{\phi_*}{\phi_e}.
\end{equation}
In the large-field limit (described as section \ref{non}), for Palatini Higgs inflation with non-minimal coupling, $n_s$, $r$ and $\alpha$ can be found using eq. \eqref{redefine} together with eqs. \eqref{strong}, \eqref{nsralpha1} and \eqref{efold1} in terms of $N_*$ 
\begin{equation}\label{higgsns}
n_s\approx1-\frac{2}{N_*}, \qquad r\approx \frac{2}{\xi N^2_*}, \qquad \alpha\approx-\frac{2}{N_*^2}.
\end{equation}
On the other hand, in the case of $\phi\ll v$ when cosmological scales exit the horizon, the potential approximates to the hilltop potential type  (described as section \ref{hilltop}) effectively
\begin{equation}\label{higgsns1}
V_E(\phi)\approx A \left[1-2\left(\frac{\phi}{v}\right)^2\right].
\end{equation}
Predictions of this potential type in eq. \eqref{higgsns1} for $\phi\ll v$ that $r$ is very suppressed and $n_s\approx1-8/v^2$. In this section, we analyze numerically for $\phi>v$ and $\phi<v$ cases in the high-$N$ scenario and low-$N$ scenario for Higgs potential with non-minimal coupling in the Palatini approach with broad range of $\xi$ and $v$. In literature, inflationary predictions of Palatini Higgs inflation taken into account for different $N_*$ values, in general taken to be constant between $N_*\approx 50-60$ \cite{Almeida:2018oid,Tenkanen:2019jiq,Takahashi:2018brt,Rubio:2019ypq,Enckell:2018kkc}. For example, \cite{Rubio:2019ypq} analyzed preheating stage following at the end of Palatini Higgs inflation taking $N_*\approx 50$. They showed that slow decaying oscillations of Higgs afterwards the end of inflation permits the field to periodically return to the plateau of the potential so the prehating stage in the Palatini Higgs inflation necessarily instantaneous. Therefore, this decreases $N_*$ of inflation required to solve the problems of hot big bang.

\begin{figure}[h!]
	\centering
	\includegraphics[angle=0, width=17cm]{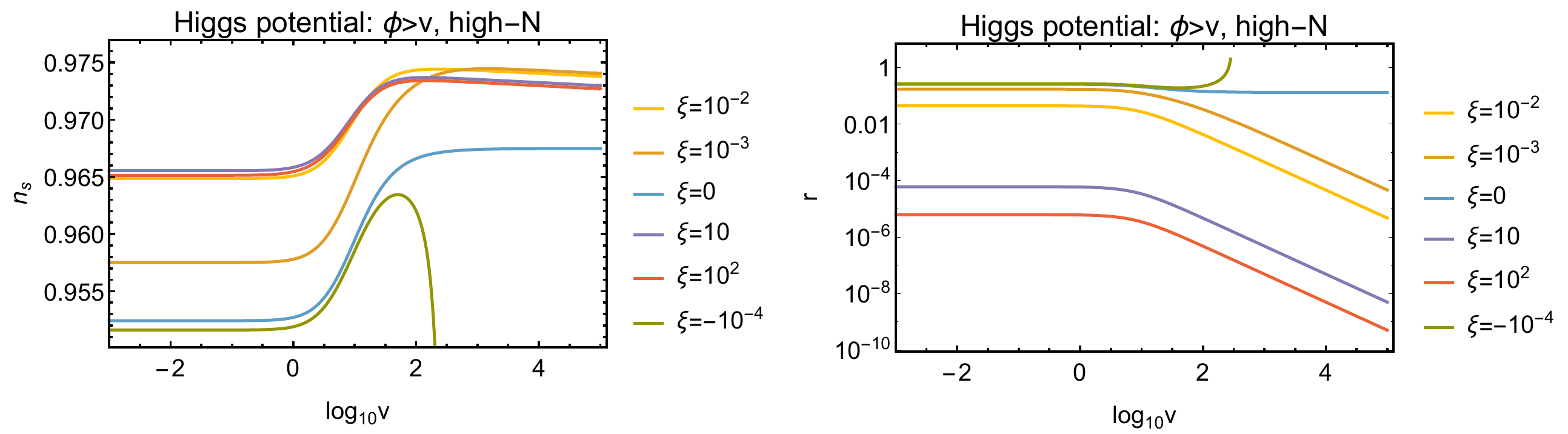}
	\includegraphics[angle=0, width=10cm]{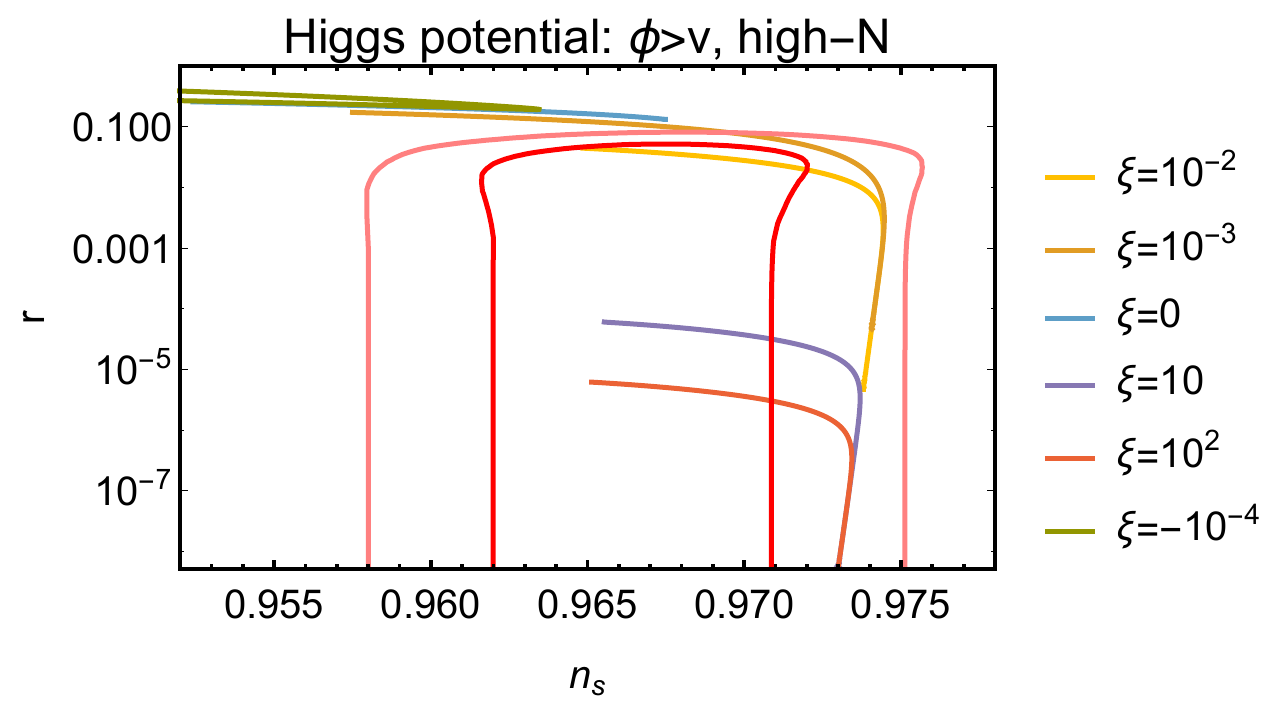}
	\caption{For Higgs potential in the Palatini formalism in the cases of $\phi>v$ and high-$N$ scenario, in the top figures display changing $n_s$ and $r$ values for different $\xi$ cases as function of $v$ and the bottom figure shows that $n_s-r$ predictions for selected $\xi$ values. The pink (red) contour correspond to the 95\% (68\%) CL contour given by the Keck Array/BICEP2 and Planck collaborations \cite{Ade:2018gkx}. }
	\label{fig6}
\end{figure}

\begin{figure}[h!]
	\centering
	\includegraphics[angle=0, width=17cm]{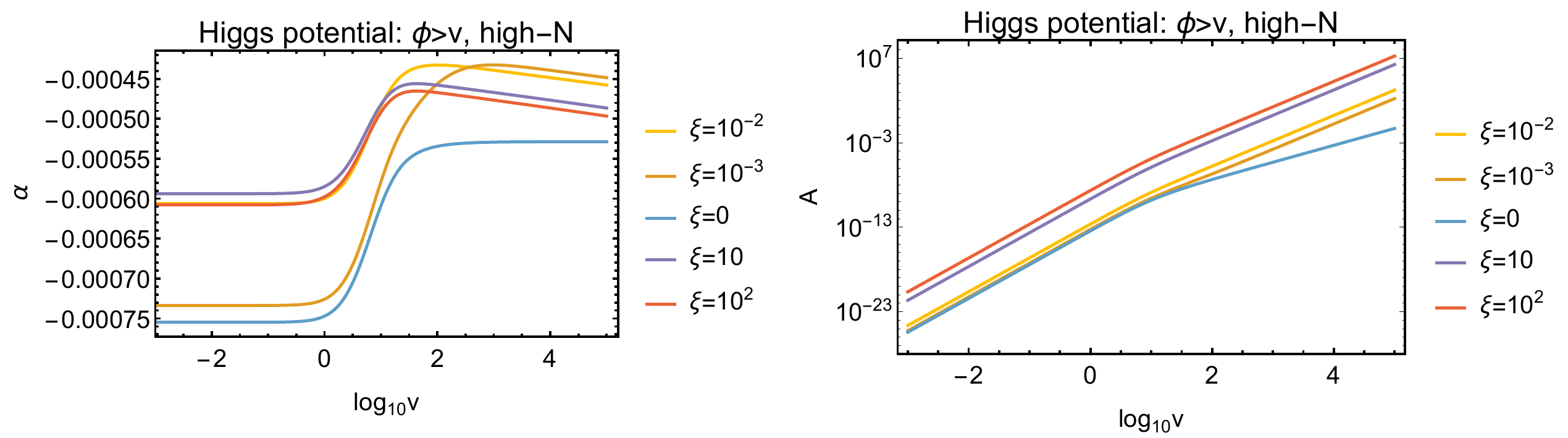}
	\caption{For Higgs potential in the Palatini formalism, the change in $\alpha$ and $A$ as a function
		of $v$ is plotted for different $\xi$ values in the cases of $\phi>v$ and high-$N$ scenario.}
	\label{fig7}
	\end{figure}

First of all, we illustrate $\phi>v$ case for both two scenarios. As it can be seen in figures \ref{fig6} and \ref{fig7}, $\xi\leq0$ cases are outside 95\% CL contour given by Keck Array/BICEP2 and Planck collaborations \cite{Ade:2018gkx} at any $v$ values. In addition to this, for small $v$ values, inflationary predictions of $\xi=10^{-3}$ can be outside 95\% CL contour. However, in larger $v$ values, predictions are inside 95\% CL for $\xi=10^{-3}$. For $\xi=10^{-2}$, predictions are inside 68\% CL for small $v$ values but for larger values of $v$, predictions remain 95\% CL contour. Furthermore, for $\xi\gg1$ cases, predictions are inside 68\% CL for small values of $v$ and when $v$ increases, they enter in the 95\% CL and $r$ is very tiny for larger and smaller values of $v$, so $r$ is highly suppressed at any $v$ values for larger $\xi$ cases. For $\xi=10^{-2}$ and $\xi=10^{-3}$, also $r$ is very small for large $v$ values but this case is not valid for smaller values of $v$. For both $\xi<0$ and $\xi=0$ cases, $r$ does not take very small values for larger and smaller $v$. In addition to this, as it can be seen that from fig. \ref{fig7}, $\alpha$ takes very tiny values for selected $\xi$ cases and at any $v$ values to be observed in the near future observations. 
Moreover, for all selected $\xi$ values, when $v$ increases, values of $A$ increase depending on $v$.

\begin{figure}[h!] 
	\centering
	\includegraphics[angle=0, width=17cm]{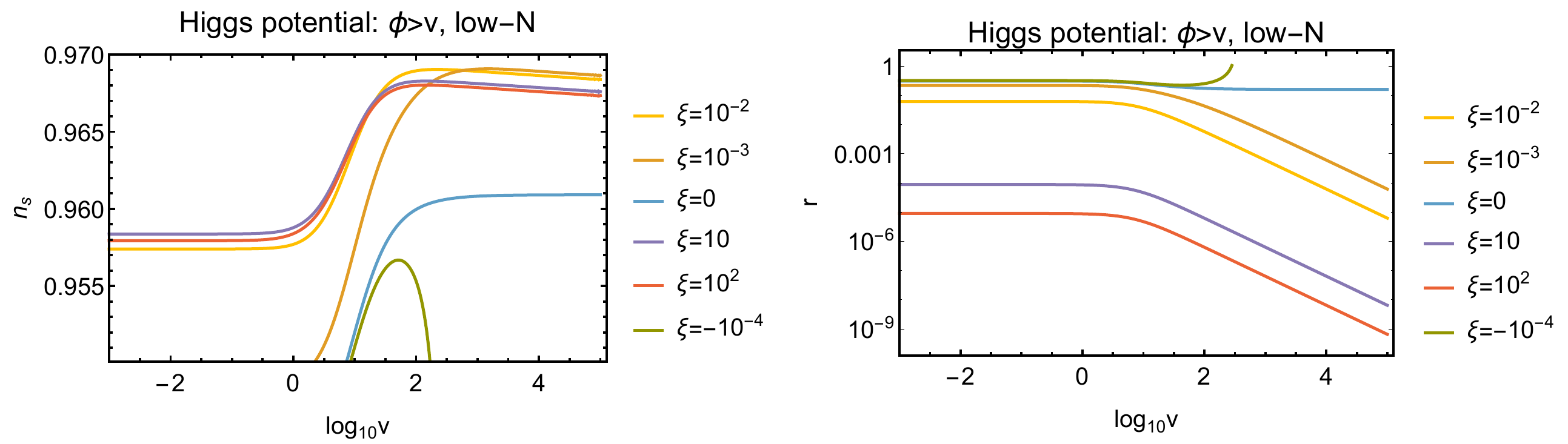}
	\includegraphics[angle=0, width=10cm]{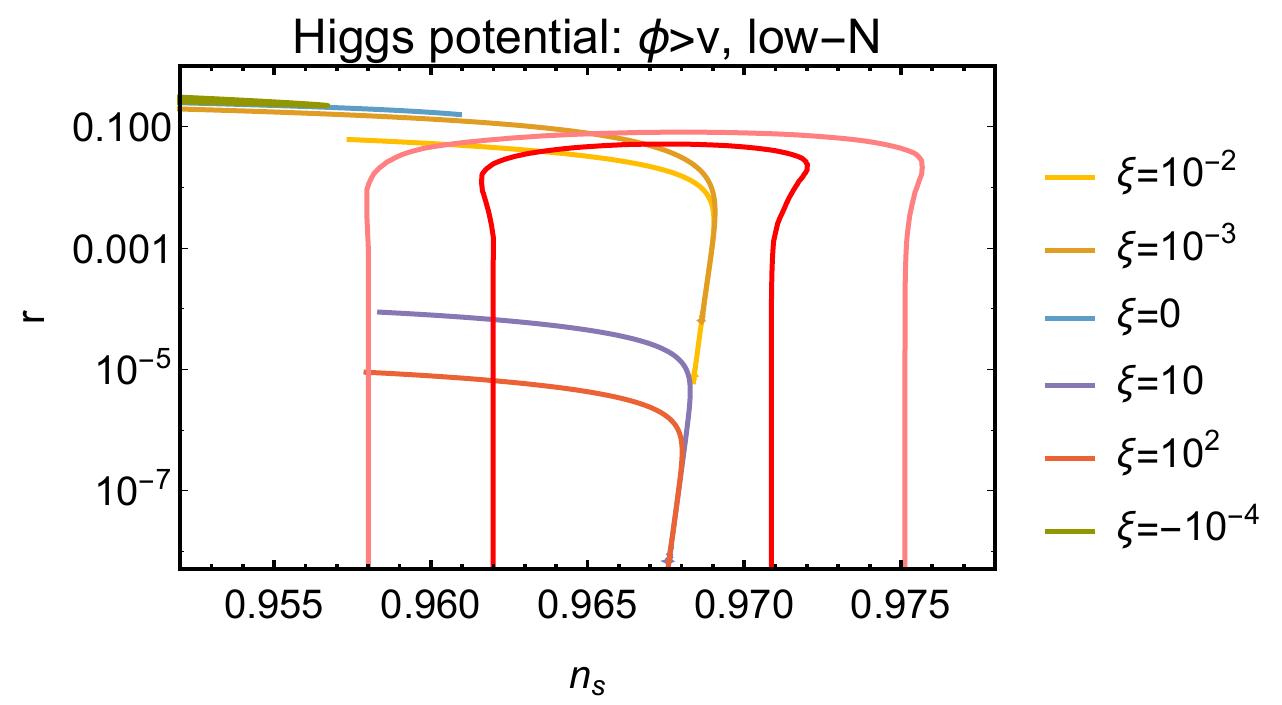}
	\caption{For Higgs potential in the Palatini formalism in the cases of $\phi>v$ and low-$N$ scenario, in the top figures display changing $n_s$ and $r$ values for different $\xi$ cases as function of $v$ and the bottom figure shows that $n_s-r$ predictions for selected $\xi$ values. The pink (red) contour correspond to the 95\% (68\%) CL contour given by the Keck Array/BICEP2 and Planck collaborations \cite{Ade:2018gkx}.}
	\label{fig8}
	
\end{figure}

\begin{figure}[h!]
	\centering
	\includegraphics[angle=0, width=17cm]{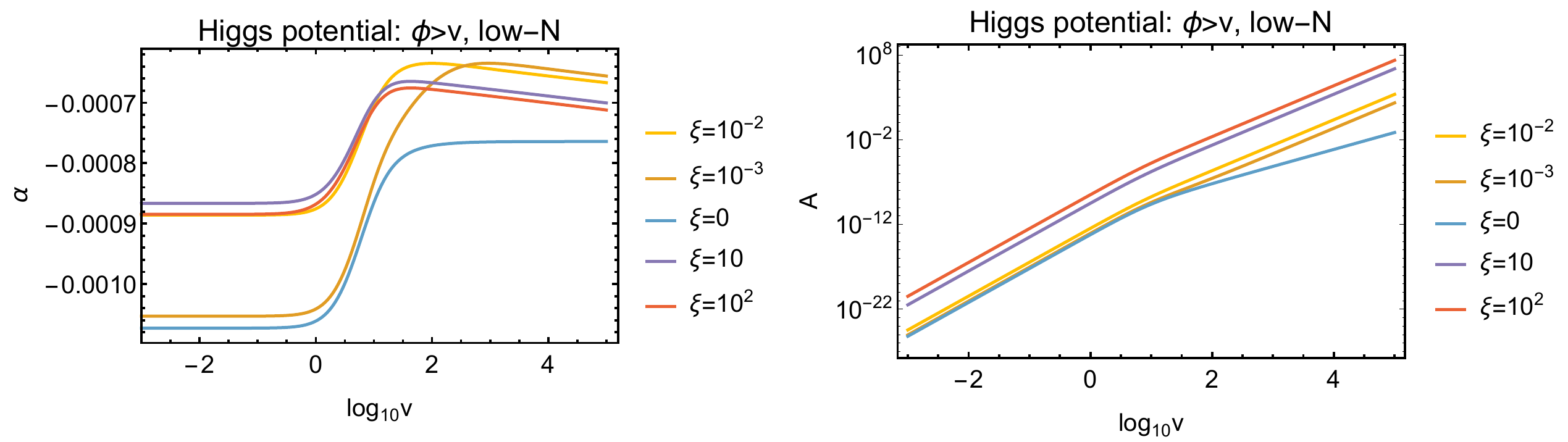}
	\caption{For Higgs potential in the Palatini formalism, the change in $\alpha$ and $A$ as a function
		of $v$ is plotted for different $\xi$ values in the cases of $\phi>v$ and low-$N$ scenario.}
	\label{fig9}
	
\end{figure}

In addition to $\phi>v$ and high-$N$ cases, figs. \ref{fig8} and \ref{fig9} show that for $\phi>v$ but low-$N$ case for Higgs potential in the Palatini formalism. According to fig. \ref{fig8}, predictions of $\xi=0$ and $\xi<0$ cases are similar as $\phi>v$ and high-$N$ scenario results. On the other hand, predictions of another $\xi$ values slightly different from high-$N$ case. For $\xi\gg1$ cases, predictions can be in the 95\% CL contour for small $v$ values but when $v$ increases, predictions remain inside 68\% CL. In the low-$N$ case, values of $r$ for all the selected $\xi$ values overlap with high-$N$ case so again $r$ is very small for $\xi\gg1$ cases for both small and large $v$ values and also for larger values of $v$ for $\xi=10^{-2}$ and $\xi=10^{-3}$ cases also $r$ is very tiny except for small $v$ values. Furthermore, in the low-$N$ case, values of $\alpha$ and $A$ are similar as high-$N$ case so  $\alpha$ takes very small values for our selected $\xi$ cases and at any $v$ values to be observed in the near future measurements.

\begin{figure}[h!]
	\centering
	\includegraphics[angle=0, width=17cm]{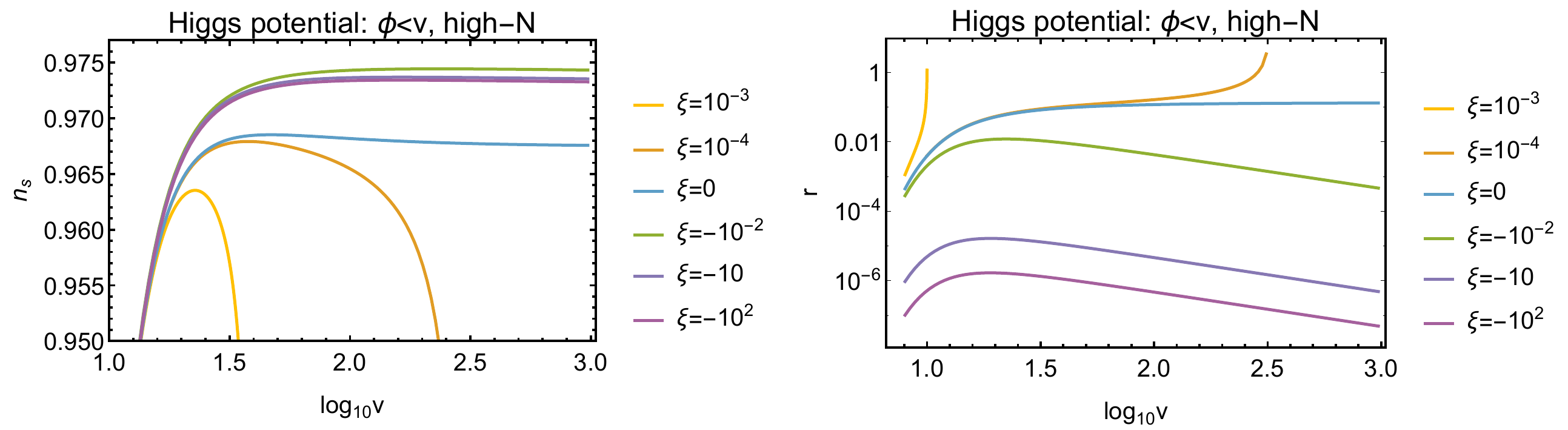}
	\includegraphics[angle=0, width=17cm]{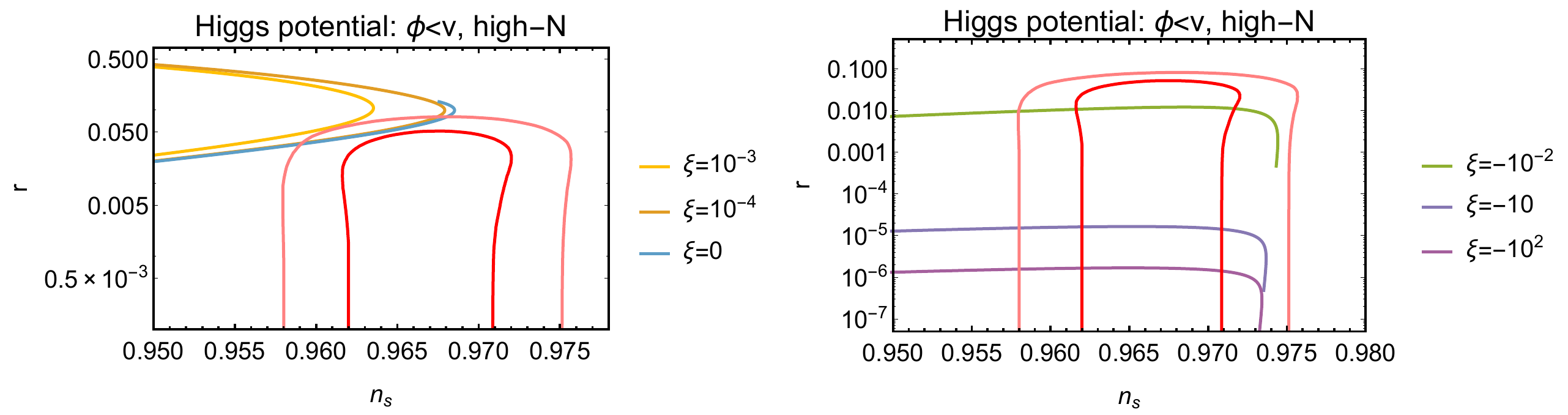}
	\caption{For Higgs potential in the Palatini formalism in the cases of $\phi<v$ and high-$N$ scenario, in the top figures display changing $n_s$ and $r$ values for different $\xi$ cases as function of $v$. The bottom figures show that $n_s-r$ predictions for selected $\xi$ values, left panel: $\xi>0$ and $\xi=0$ cases, right panel: $\xi<0$ cases. The pink (red) contour correspond to the 95\% (68\%) CL contour given by the Keck Array/BICEP2 and Planck collaborations \cite{Ade:2018gkx}.}
	\label{fig10}
\end{figure}

\begin{figure}[h!]
	\centering
	\includegraphics[angle=0, width=17cm]{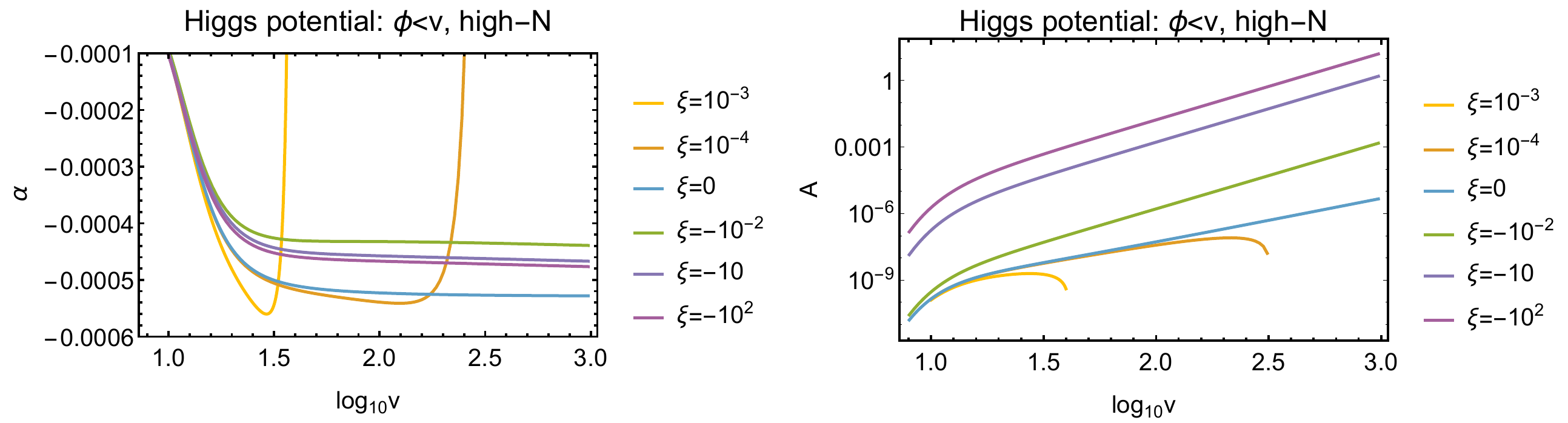}
	
	\caption{For Higgs potential in the Palatini formalism, the change in $\alpha$ and $A$ as a function
		of $v$ is plotted for different $\xi$ values in the cases of $\phi<v$ and high-$N$ scenario.}
	\label{fig11}
	
\end{figure}

Numerical results for $\phi<v$ and high-$N$ cases for Higgs potential in the Palatini approach can be seen in figures \ref{fig10} and \ref{fig11}, according to these figures, predictons of $\xi=10^{-3}$ are ruled out for current data. In contrast, $\xi=10^{-4}$ and $\xi=0$ cases can be inside 95\% CL contour given by the Keck Array/BICEP2 and Planck collaborations \cite{Ade:2018gkx}. However, $\xi<0$ cases for the range between $10\lesssim v \lesssim20$, predictions are outside 95\% CL contour but when $v$ increases, they can be in the range of compatible with observational data depending on $v$. Unlike from $\phi>v$ and high-$N$ scenario, here values of $r$ are very small for $\xi=-10$ and $\xi=-10^2$ cases. In addition to this, $\alpha$ values are very small similar to other situations. Lastly, for $\xi\leqslant0$ cases, values of $A$ increase depending on $v$, but this case is different for $\xi=10^{-3}$ and $\xi=10^{-4}$ values, as it can be seen in fig. \ref{fig11}.

\begin{figure}[h!]
	\centering
	
	\includegraphics[angle=0, width=17cm]{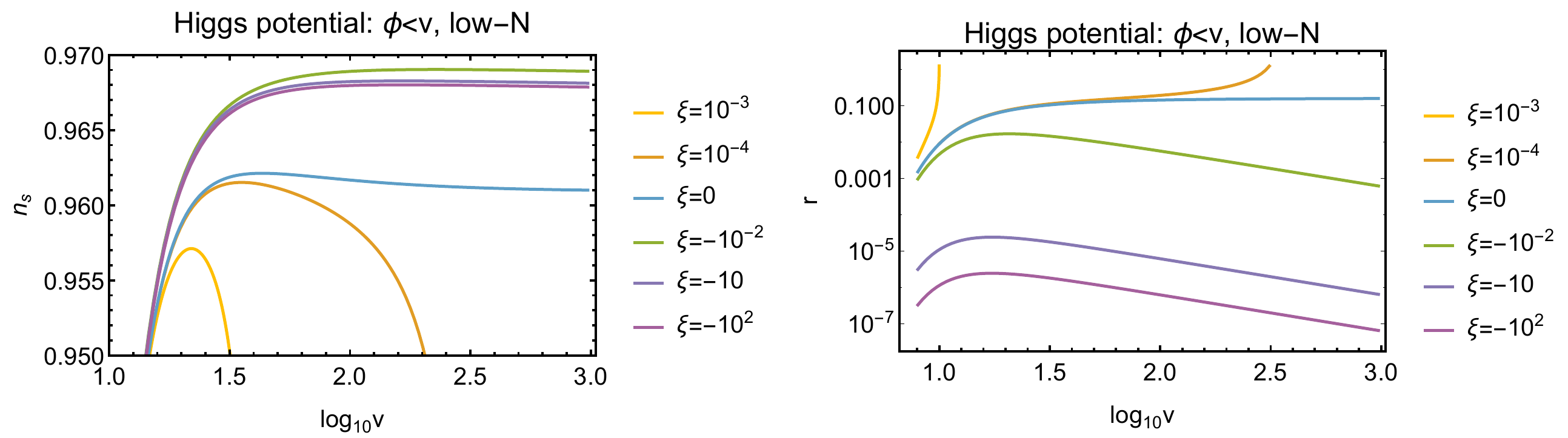}
	
	\includegraphics[angle=0, width=17cm]{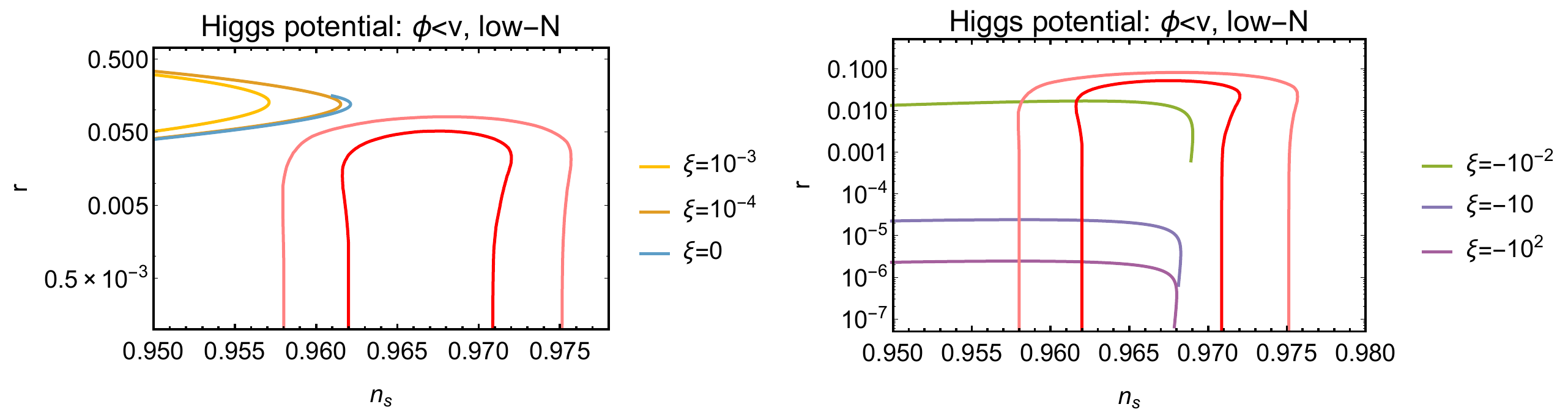}
	\caption{For Higgs potential in the Palatini formalism in the cases of $\phi<v$ and low-$N$ scenario, in the top figures display changing $n_s$ and $r$ values for different $\xi$ cases as function of $v$. The bottom figures show that $n_s-r$ predictions for selected $\xi$ values, left panel: $\xi>0$ and $\xi=0$ cases, right panel: $\xi<0$ cases. The pink (red) contour correspond to the 95\% (68\%) CL contour given by the Keck Array/BICEP2 and Planck collaborations \cite{Ade:2018gkx}.}
	\label{fig12}
	
\end{figure}

\begin{figure}[h!]
	\centering
	\includegraphics[angle=0, width=17cm]{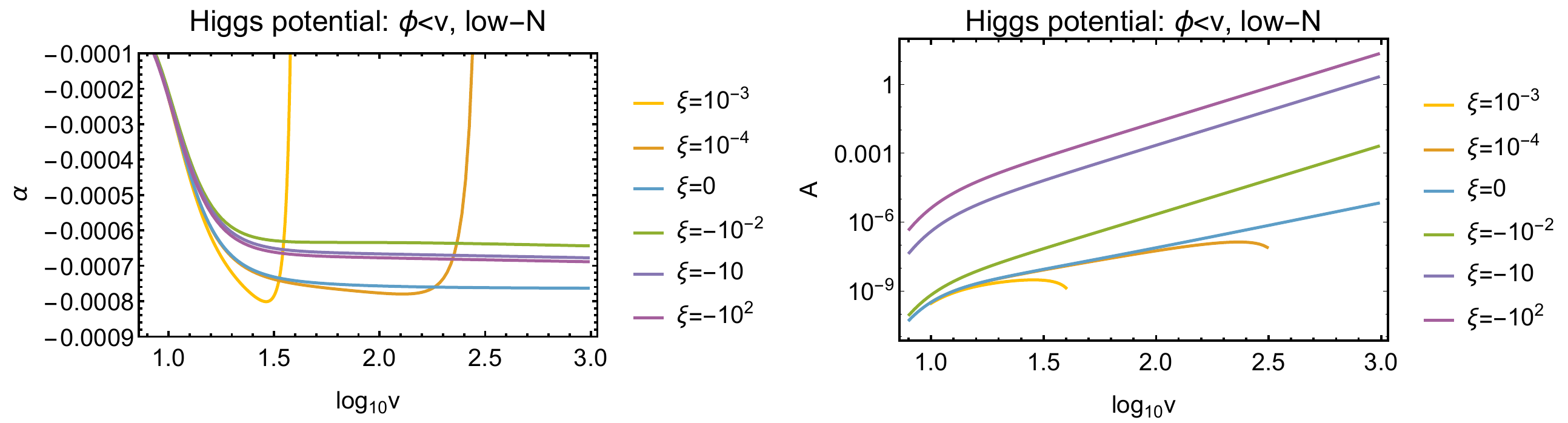}
	\caption{For Higgs potential in the Palatini formalism, the change in $\alpha$ and $A$ as a function of $v$ is plotted for different $\xi$ values in the cases of $\phi<v$ and low-$N$ scenario.}
	\label{fig13}
\end{figure}

\begin{figure}[h!]
	\centering
	\includegraphics[angle=0, width=17cm]{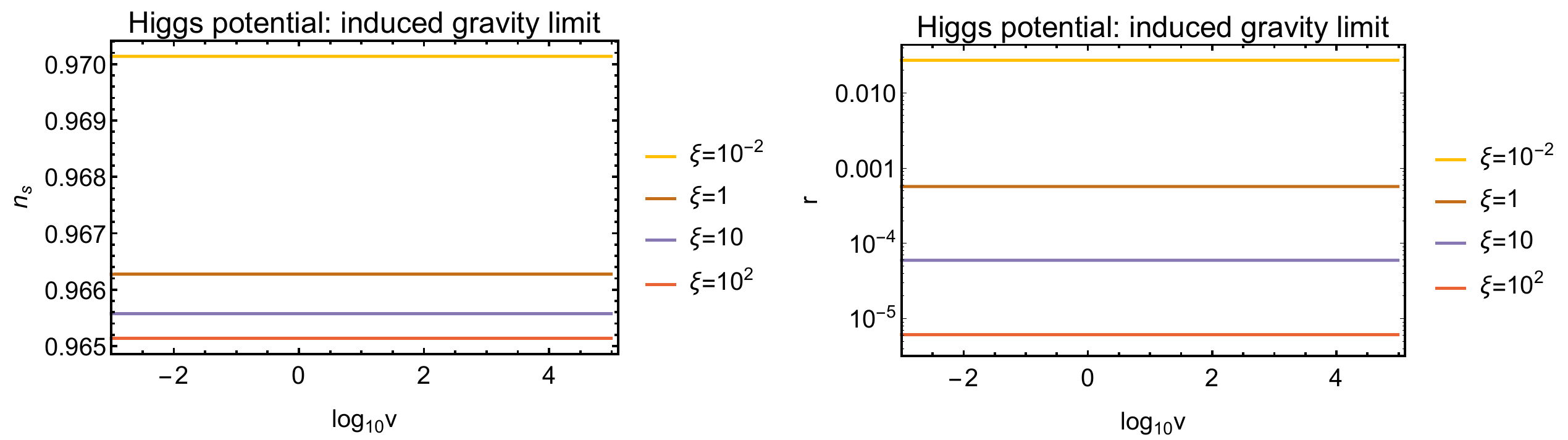}
	\includegraphics[angle=0, width=17cm]{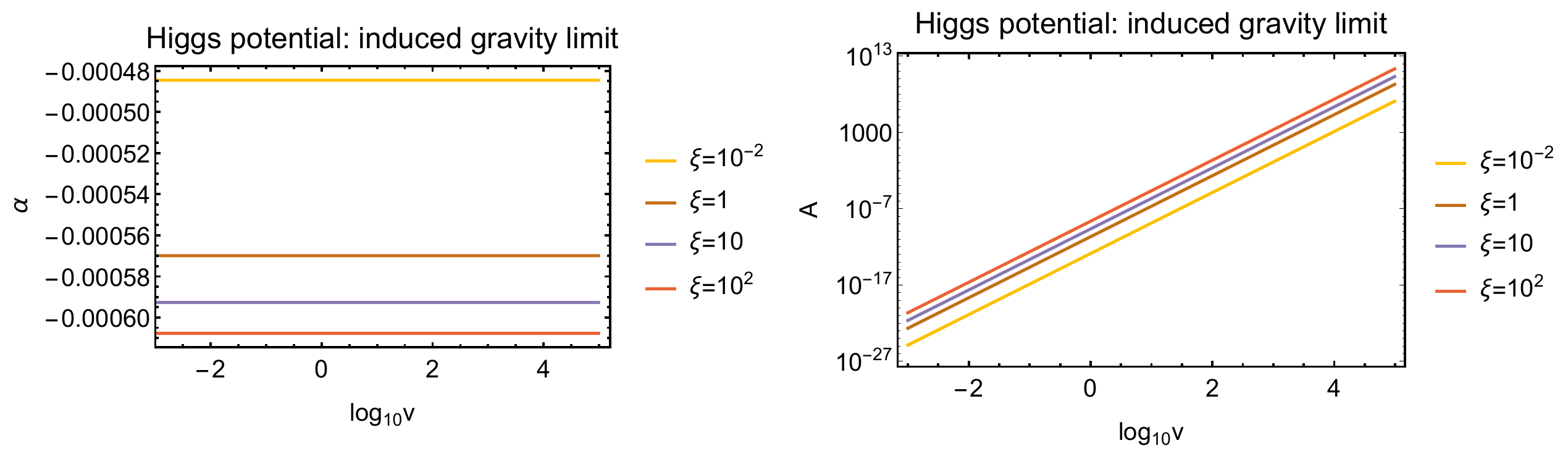}
	\caption{For Higgs potential in the Palatini formalism, the change in $n_s$, $r$, $\alpha$ and $A$ as a function
		of $v$ is plotted for different $\xi$ values and for $\phi>v$ in the induced gravity limit which corresponds to $\xi v^2=1$.}
	\label{fig14}
\end{figure}

Furthermore, we also obtain numerical results in the cases of $\phi<v$ and low-$N$ scenario for Higgs potential in figures \ref{fig12} and \ref{fig13}. According to these figures, inflationary predictions of $\xi\geq0$ cases are ruled out for current data. In addition to this, the cases of $\xi<0$, predictions begin to enter observational region, when $v$ increases. For larger $v$ values, predictions remain 68\% CL contour, as well as values of $r$ are strongly suppressed for cases of both $\xi=-10$ and $\xi=-10^2$. According to fig. \ref{fig13}, results for the $\alpha$ and $A$ are the same as $\phi<v$ and high-$N$ scenario. We also display inflationary parameters of Higgs potential in the limit of induced gravity, described as in the text (see section \ref{non}) for high-$N$ scenario in fig. \ref{fig14}. According to this figure, for all our selected $\xi$ values are in the  68\% CL contour. What is more, in this limit case, $\alpha$ values are also very tiny at any $v$ and values of $A$ increase, depending upon $v$. 

In the induced gravity limit, using eq. \eqref{induced}, the Einstein frame potential can be obtained in terms of $\chi$ in that form
\begin{equation}\label{inducedeq}
V_E(\chi)=\frac{A}{\xi^2 v^4}\left(1-2\exp\left(-2\sqrt{\xi}\chi\right)\right).
\end{equation}

For this potential, using eq. \eqref{efold1}, $8\xi N \approx \exp\left(2\sqrt{\xi}\chi\right)$. Therefore, using eq. \eqref{nsralpha1} we can find $n_s$ and $r$ approximately in the induced gravity limit
\begin{equation}\label{inducedeq1}
n_s\approx1-\frac{2}{N_*}-\frac{3}{4\xi N^2_*}, \qquad r\approx \frac{2}{\xi N_*^2}.
\end{equation}
The Higgs potential in the induced gravity limit was previously investigated for the Metric formulation in \cite{Burns:2016ric,Kaiser:1994vs,Cerioni:2009kn}. In this work, we extend these papers with analyzing Higgs potential in the induced gravity limit in the Palatini formulation for $\phi>v$ and high-$N$ scenario. To sum up, in literature, Higgs inflation with non-minimal coupling has been discussed such refs. \cite{Linde:2011nh,Bostan:2018evz,Burns:2016ric,Kaiser:1994vs,Cerioni:2009kn,Tronconi:2017wps} in the Metric formulation. Refs. \cite{Linde:2011nh} and (for just $\xi>0$) \cite{Tronconi:2017wps} analyzed the Higgs inflation with non-minimal coupling in the Metric formulation in general for taking $F(\phi)=1+\xi \phi^2$. Moreover, \cite{Bostan:2018evz} explained the Higgs inflation with non-minimal coupling in the Metric formulation for both $\xi>0$ and $\xi<0$ cases for $F(\phi)=1+\xi(\phi^2-v^2)$. On the other hand, some papers take into account non-minimally coupled Higgs inflation in the Palatini formulation \cite{Bauer:2008zj,Rasanen:2017ivk,Jinno:2019und,Rubio:2019ypq} which we mentioned before. Ref. \cite{Bauer:2008zj} examined for large-field limit taking $F(\phi)=1+\xi \phi^2$ and they found $n_s\simeq0.968$ ad $r\simeq10^{-14}$ in the Palatini approach. Moreover, \cite{Rasanen:2017ivk} again taking $F(\phi)=1+\xi \phi^2$, they found predictions of various inflationary parameters in Palatini approach, they obtained that $r$ values are highly suppressed for $\xi\phi^2\gg1$ limits and also they found very small $\alpha$ values to be observed in the future measurements. Similar to the other papers, \cite{Rubio:2019ypq} analyzed Palatini Higgs inflation taking $F(\phi)=1+\xi \phi^2$. Different from previous papers, we analyze inflationary parameters of Palatini Higgs inflation with non-minimal coupling for taking $F(\phi)=m^2+\xi \phi^2=1+\xi (\phi^2-v^2)$. Furthermore, we display our numerical calculations using both high-$N$ scenario and low-$N$ scenario.  
\section{Hilltop potentials}\label{hilltop}
\begin{figure}[h!]
	\centering
	\includegraphics[angle=0, width=16cm]{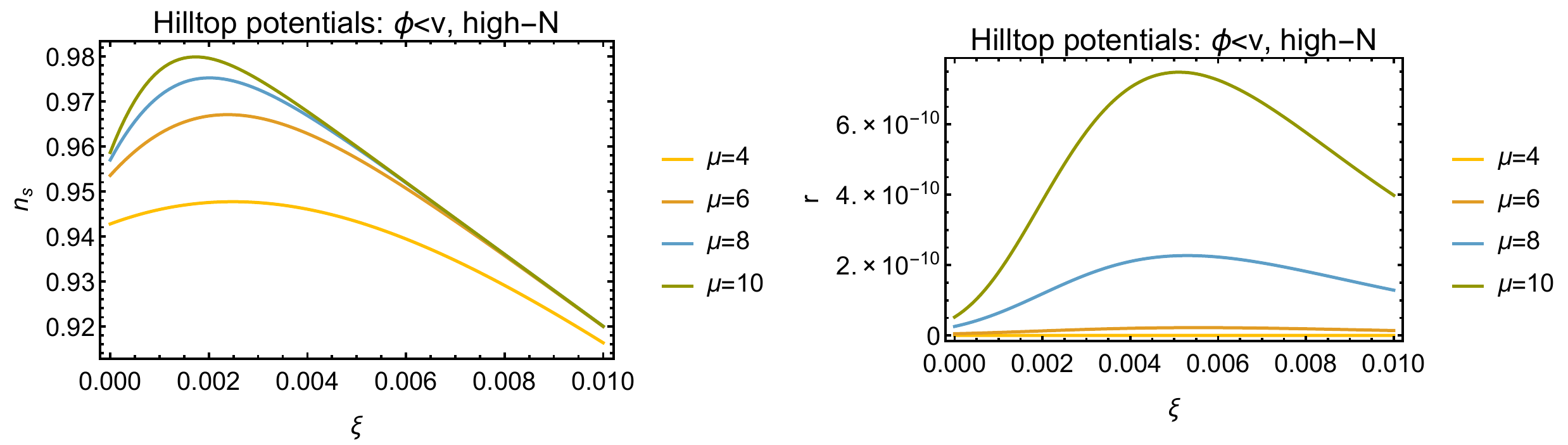}
	\includegraphics[angle=0, width=9cm]{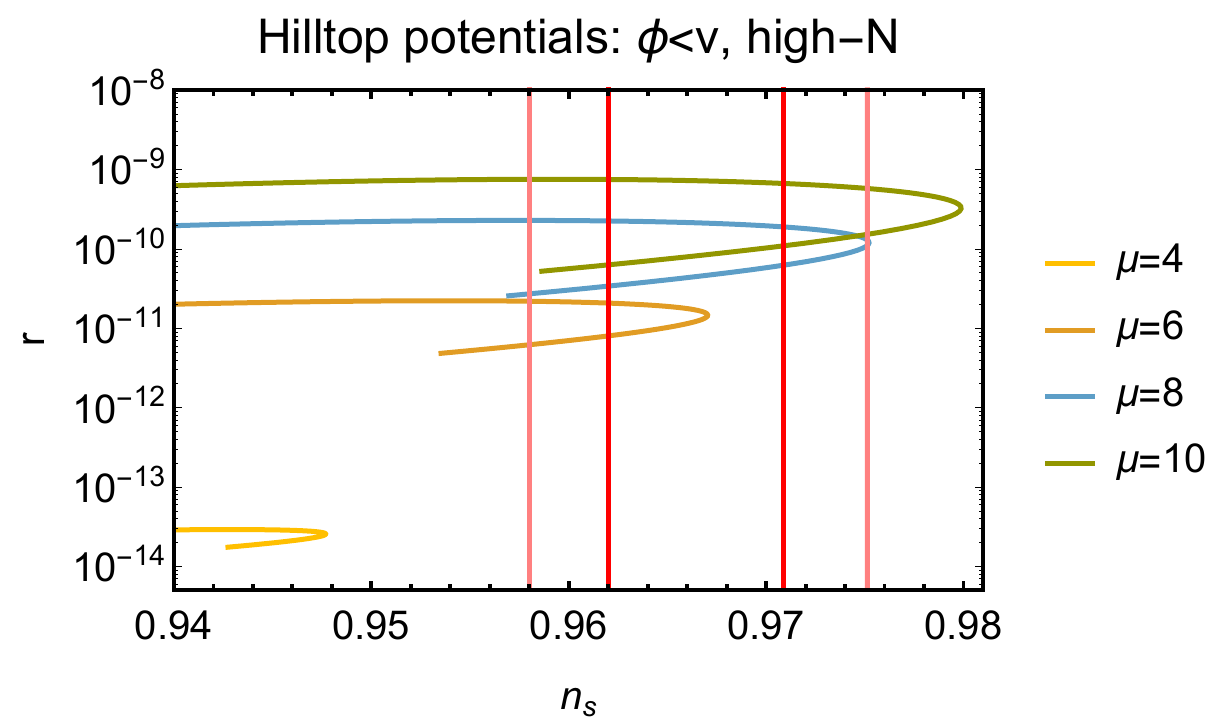}
	
	\caption{For hilltop potentials in the Palatini formalism, top figures display that $n_s$, $r$ values as functions of $\xi$ for $\tau=0.01$ and different $\mu$
		values in the cases of $\phi<v$ and high-$N$ scenario. The bottom figure displays $n_s-r$ predictions based on range of the top figures $\xi$ values for $\tau=0.01$. The pink
		(red) line corresponds to the 95\% (68\%) CL contour
		given by the Keck Array/BICEP2 and Planck collaborations \cite{Ade:2018gkx}.}
	\label{fig15}
	\end{figure}

\begin{figure}[h!]
	\centering
	\includegraphics[angle=0, width=9cm]{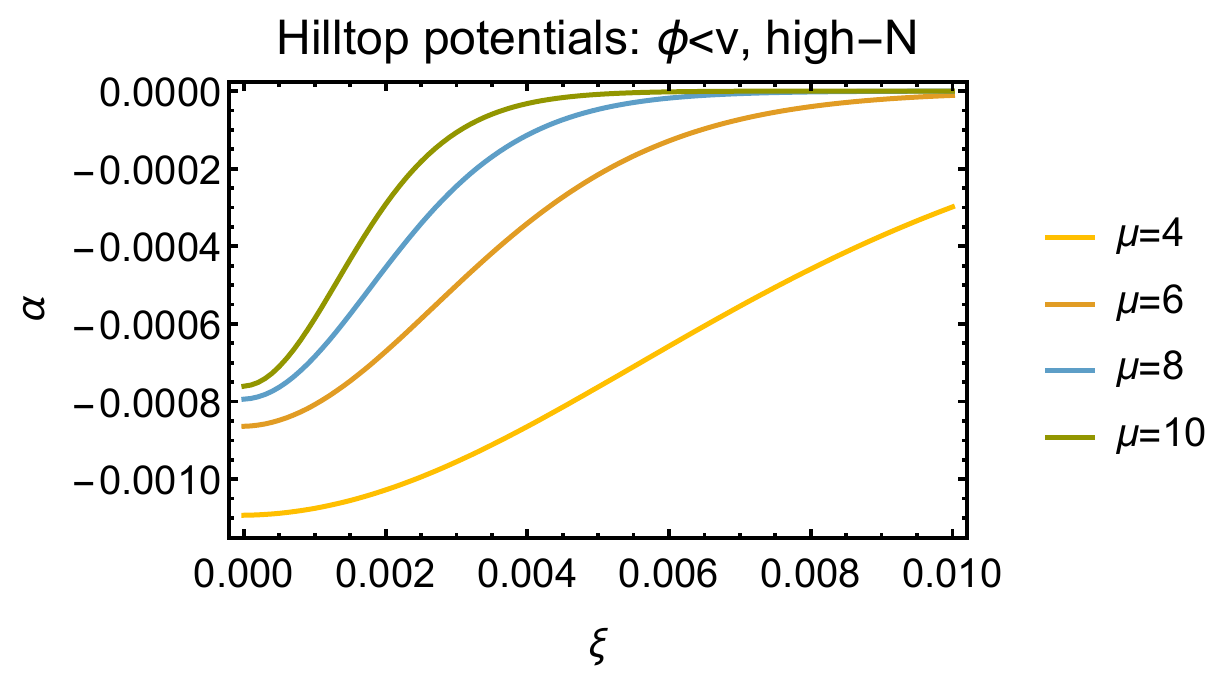}
	\caption{For hilltop potentials in the Palatini formalism, the figure shows that $\alpha$ values as functions of $\xi$ for $\tau=0.01$ and different $\mu$
		values in the cases of $\phi<v$ and high-$N$ scenario.}
	\label{fig16}
	
\end{figure}

\begin{figure}[h!]
	\centering
	\includegraphics[angle=0, width=16cm]{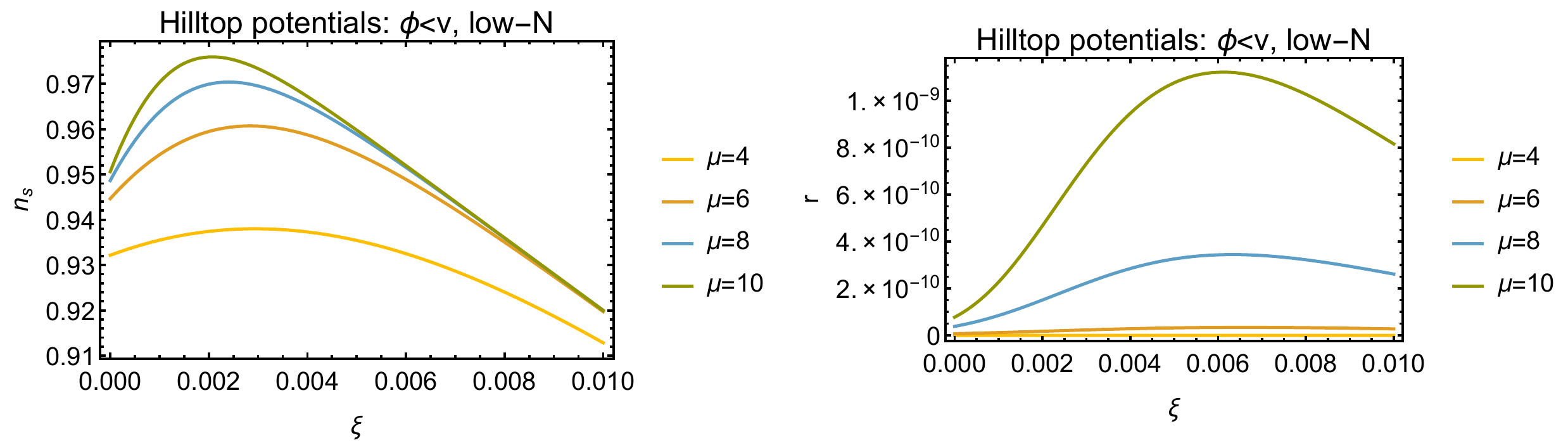}
	\includegraphics[angle=0, width=9cm]{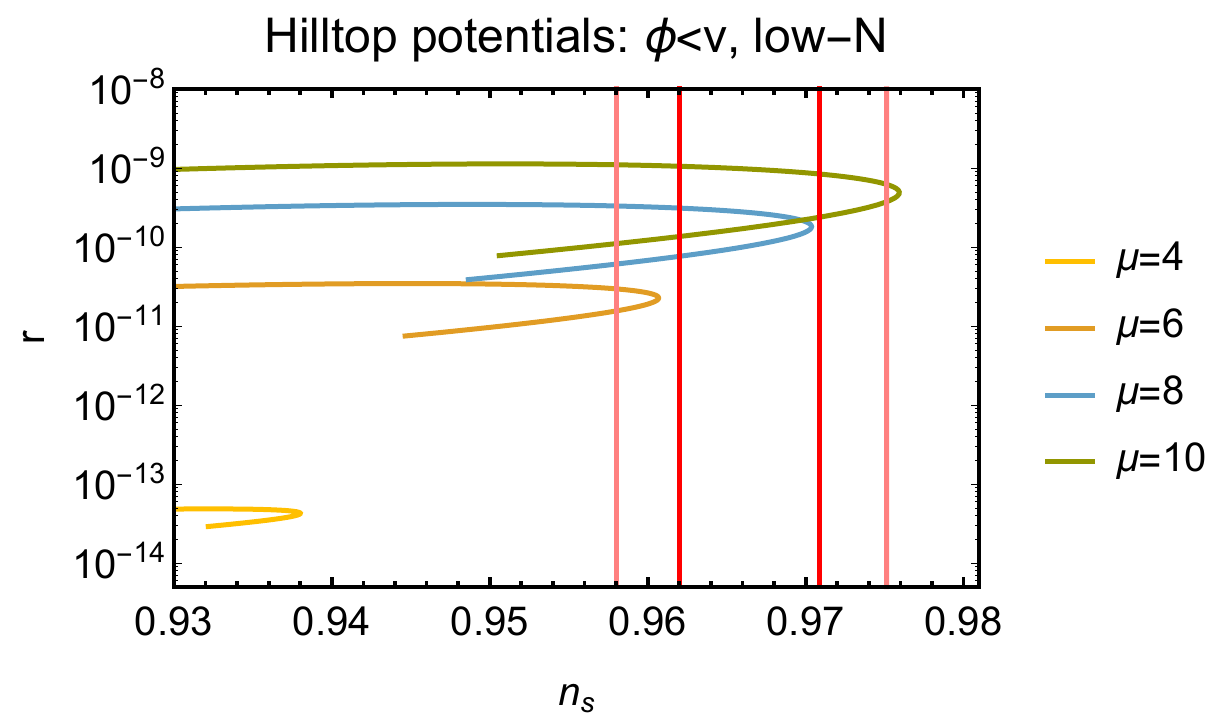}
	
	\caption{For hilltop potentials in the Palatini formalism, top figures display that $n_s$, $r$ values as functions of $\xi$ for $\tau=0.01$ and different $\mu$
		values in the cases of $\phi<v$ and low-$N$ scenario. The bottom figure displays $n_s-r$ predictions based on range of the top figures $\xi$ values for $\tau=0.01$. The pink
		(red) line corresponds to the 95\% (68\%) CL contour
		given by the Keck Array/BICEP2 and Planck collaborations \cite{Ade:2018gkx}.}
	\label{fig17}
	\end{figure}

\begin{figure}[h!]
	\centering
	\includegraphics[angle=0, width=9cm]{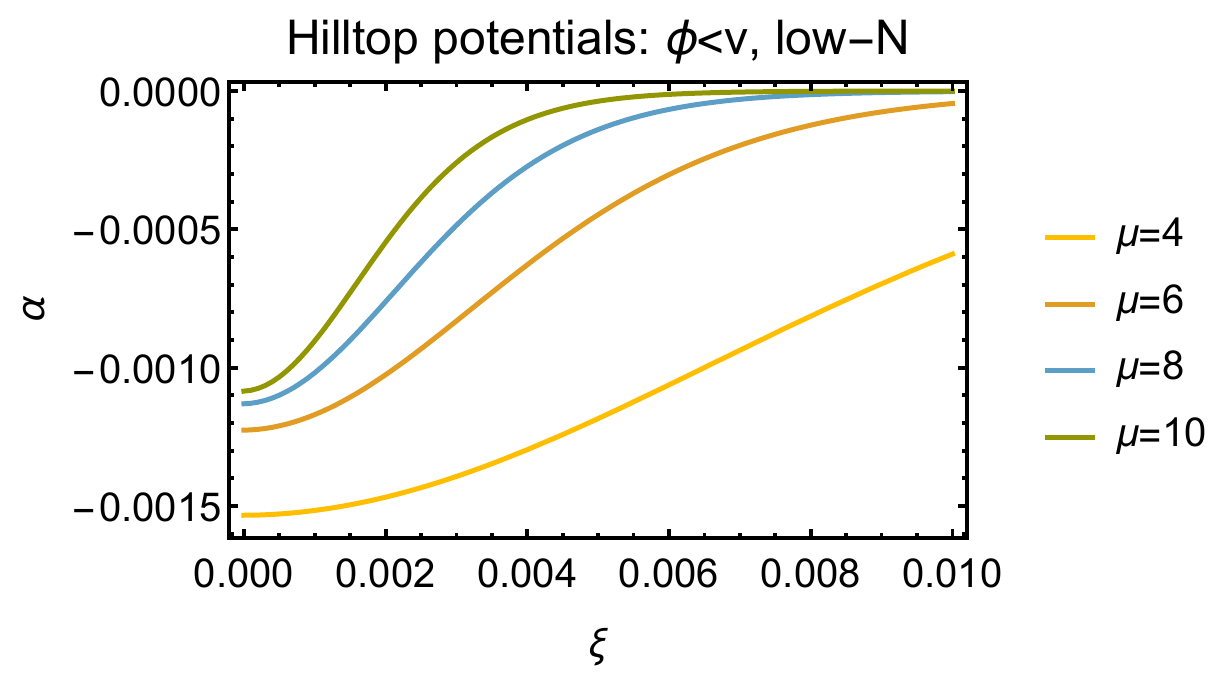}
	\caption{For hilltop potentials in the Palatini formalism, the figure shows that $\alpha$ values as functions of $\xi$ for $\tau=0.01$ and different $\mu$
		values in the cases of $\phi<v$ and low-$N$ scenario.}
	\label{fig18}
\end{figure}

In this section, we take into account another symmetry-breaking type potential models which also take place in some supersymmetric inflation models, i.e. \cite{Izawa:1996dv,Kawasaki:2003zv,Senoguz:2004ky} in the case of the inflaton value is $\phi<v$ throughout inflation. These potential types are described with generalization of the Higgs potential in that form
\begin{equation}\label{generalizedhiggs}
V_J(\phi)=A\left[1-\left(\frac{\phi}{v}\right)^\mu\right]^2\,,\quad(\mu>2)\,.
\end{equation}
In the electroweak regime which explained as section \ref{non}, we have $\phi\approx\chi$ and also $\chi\ll v$ during inflation, and the Einstein frame potential can be obtained as in terms of canonical scalar field
\begin{equation}\label{sfipotential}
V_E(\chi)\approx
A\left[1-\left(\frac{\chi}{\tau}\right)^\mu-2\xi\chi^2\right]\,,
\end{equation}
where we have defined $\tau=v/2^{1/\mu}$. In the literature, hilltop potentials with minimal coupling case ($\xi=0$) has been investigated such refs. \cite{Lyth:2009zz,Martin:2013tda}. Furthermore, by taking consideration \eqref{slowroll1}, \eqref{nsralpha1} and \eqref{efold1}, we find 
\begin{equation}\label{nsrminimal}
n_s\approx1-\frac{(\mu-1)2}{(\mu-2)N_*}\,,\quad
r\approx128\left(\frac{16\tau^{2\mu}}{\mu^2[4\mu-2)N_*]^{2\mu-2}}\right)^\frac{1}{\mu-2}\,,
\end{equation}
which illustrates that $r$ is strongly suppressed and $n_s$ takes to be smaller values than the range agreement with observational results. On the other hand, in this work we calculate inflationary parameters for hilltop potentials with non-minimal coupling in Palatini formulation both high-$N$ scenario and low-$N$ scenario numerically. In these calculations are shown in figures \ref{fig15}, \ref{fig16}, \ref{fig17} and \ref{fig18}. Furthermore, for potential in eq. \eqref{sfipotential}, $n_s$ and $r$ can be obtained in that form
\begin{equation}\label{sfinsr}
n_s\approx1+\frac{8(\mu-1)\xi}{1-e^{4(\mu-2)\xi N_*}}-8\xi\,,\quad
r\approx\frac{128\xi^2\tau^2(4\xi\tau^2/\mu)^{2/(\mu-2)}e^{8(\mu-2)\xi N_*}}{
	\left(e^{4(\mu-2)\xi N_*}-1\right)^{2(\mu-1)/(\mu-2)}}\,.
\end{equation}
These predictions are compatible with the our numerical results for $n_s-r$ that were computed using the Jordan frame potential described by eq. \eqref{generalizedhiggs} which is shown top figures in the fig. \ref{fig15} and fig. \ref{fig17} for two different scenarios. As it can be seen that figures \ref{fig15}, \ref{fig16}, \ref{fig17} and \ref{fig18} in general, on the condition that $\xi,v\ll1$ and $\tau=0.01$, observational parameters can be inside observational region except for $\mu=4$ since predictions are ruled out for any $\xi$ values in the case of $\mu=4$ for both two scenarios which we take into account. In addition to this, as the $\xi$ values increase, observational parameters are ruled out for any $\xi$ values in the cases of $\mu=6,8,10$ different from smaller $\xi$ values as well as $r$ values are highly suppressed and also values of $\alpha$ are very tiny to be observed in the near future measurements for all selected $\mu$ values.
\clearpage
\section{Conclusion}\label{conc}
In this work, we briefly expressed Palatini inflation with a non-minimal coupling in section \ref{non} and then we displayed our results to the inflationary predictions of non-minimally coupled Palatini quadratic potential in the large-field limit for high-$N$ scenario and low-$N$ scenario in section \ref{quadraticpot} for $F(\phi)=1+\xi \phi^2$. Next, we analyzed predictions of Higgs potential for $\phi>v$ and $\phi<v$ in section \ref{higgs} and hilltop potentials for $\phi<v$ in section \ref{hilltop} with non-minimal coupling in the Palatini formulation taking $F(\phi)=1+\xi(\phi^2-v^2)$ for both two $N$ scenarios. Furthermore, in section \ref{higgs}, we also investigated Higgs potential in the induced gravity limit for high-$N$ scenario. 

We illustrated that for the Palatini quadratic potential with non-minimal coupling, just a small $\xi$ values fit the current measurements given by the Keck Array/BICEP2 and Planck collaborations \cite{Ade:2018gkx} for high-$N$ case. On the other hand, for low-$N$ case, we found that predictions are outside to the observational region for any $\xi$ values. 
According to the our results, $r$ has very tiny values in the $\xi\gg1$ cases where the inflaton value $\phi>v$ for Higgs potential for high-$N$ scenario and low-$N$ scenario. Therefore, we obtained the significant Starobinsky attractor behaviour for larger $\xi$ values in the Metric formulation is disappear in the Palatini formulation for these $\xi$ cases where the inflaton value $\phi>v$ for both two scenarios. In addition to this, for $\xi=10^{-2}$ and $\xi=10^{-3}$, $r$ has very tiny values solely larger $v$. However, in the case of $\phi<v$ and for also both two scenarios, $r$ values are highly suppressed for $\xi=-10$ and $\xi=-10^2$. 

We also analyzed Palatini Higgs inflation in the induced gravity limit for high-$N$ scenario and we found that for $\xi\geq1$ cases, $r$ takes small values. Furthermore, we calculated the inflationary predictions of hilltop potentials numerically in the case of the inflaton value $\phi<v$ and $\xi,v\ll1$ for high-$N$ scenario and low-$N$ scenario. In these type of potentials, inflationary parameters can be compatible with approximately $\xi\lesssim0.005$ values just in the cases of $\phi<v$ and $v\ll1$. We also obtained that $r$ values are highly suppressed in the hilltop potentials for both two scenarios.  

Finally, we obtained that the predict of $\alpha$ is too small to be observed in future measurements for all our examined potentials but we consider that future experimental can be much enhanced values of $\alpha$, in particular near future observations of the 21 cm line \cite{Kohri:2013mxa,Basse:2014qqa,Munoz:2016owz}.





\begin{thebibliography}{99}
\bibitem{Guth:1980zm} 
A.~H.~Guth,
\textit{The Inflationary Universe: A Possible Solution to the Horizon and Flatness Problems},
\href{https://doi.org/10.1103/PhysRevD.23.347}{Phys.\ Rev.\ D {\bf 23}, 347 (1981)}. 

\bibitem{Linde:1981mu} 
A.~D.~Linde, \textit{A New Inflationary Universe Scenario: A Possible Solution of the Horizon, Flatness, Homogeneity, Isotropy and Primordial Monopole Problems},
\href{https://doi.org/10.1016/0370-2693(82)91219-9}{ Phys.\ Lett.\  {\bf 108B}, 389 (1982)}.  

\bibitem{Albrecht:1982wi} 
A.~Albrecht and P.~J.~Steinhardt,
\textit{Cosmology for Grand Unified Theories with Radiatively Induced Symmetry Breaking}, \href{https://doi.org/10.1103/PhysRevLett.48.1220}{Phys.\ Rev.\ Lett.\  {\bf 48}, 1220 (1982)}.



\bibitem{Linde:1983gd} 
A.~D.~Linde, \textit{Chaotic Inflation}, \href{https://doi.org/10.1016/0370-2693(83)90837-7}{Phys.\ Lett.\  {\bf 129B}, 177 (1983)}.	

\bibitem{Aghanim:2018eyx} 
N.~Aghanim {\it et al.} [Planck Collaboration],
\textit{Planck 2018 results. VI. Cosmological parameters}, 
\href{https://arxiv.org/abs/1807.06209}{1807.06209}.

\bibitem{Akrami:2018odb} 
Y.~Akrami {\it et al.} [Planck Collaboration],
\textit{Planck 2018 results. X. Constraints on inflation}, \href{https://arxiv.org/abs/1807.06211}{1807.06211}.

\bibitem{Kohri:2013mxa} 
K.~Kohri, Y.~Oyama, T.~Sekiguchi and T.~Takahashi,
\textit{Precise Measurements of Primordial Power Spectrum with 21 cm Fluctuations},
\href{https://doi.org/10.1088/1475-7516/2013/10/065}{JCAP {\bf 1310}, 065 (2013)}
[\href{https://arxiv.org/abs/arXiv:1303.1688}{1303.1688}].

\bibitem{Basse:2014qqa} 
T.~Basse, J.~Hamann, S.~Hannestad and Y.~Y.~Y.~Wong,
\textit{Getting leverage on inflation with a large photometric redshift survey},
\href{https://doi.org/10.1088/1475-7516/2015/06/042}{JCAP {\bf 1506}, no. 06, 042 (2015)}
[\href{https://arxiv.org/abs/arXiv:1409.3469}{1409.3469}].

\bibitem{Munoz:2016owz} 
J.~B.~Mu\~{n}oz, E.~D.~Kovetz, A.~Raccanelli, M.~Kamionkowski and J.~Silk,
\textit{Towards a measurement of the spectral runnings},
\href{https://doi.org/10.1088/1475-7516/2017/05/032}{JCAP {\bf 1705}, 032 (2017)}
[\href{https://arxiv.org/abs/arXiv:1611.05883}{1611.05883}].


\bibitem{Ade:2018gkx} 
P.~A.~R.~Ade {\it et al.} [BICEP2 and Keck Array Collaborations],
\textit{BICEP2 / Keck Array x: Constraints on Primordial Gravitational Waves using Planck, WMAP, and New BICEP2/Keck Observations through the 2015 Season}, \href{https://doi.org/10.1103/PhysRevLett.121.221301}{Phys.\ Rev.\ Lett.\  {\bf 121}, 221301 (2018)}
[\href{https://arxiv.org/abs/arXiv:1810.05216}{1810.05216}].

\bibitem{Wu:2016hul} 
W.~L.~K.~Wu {\it et al.},
\textit{Initial Performance of BICEP3: A Degree Angular Scale 95 GHz Band Polarimeter},
\href{https://doi.org/10.1007/s10909-015-1403-x}{J.\ Low.\ Temp.\ Phys.\  {\bf 184}, no. 3-4, 765 (2016)}
[\href{https://arxiv.org/abs/arXiv:1601.00125}{1601.00125}].

\bibitem{Matsumura:2013aja} 
T.~Matsumura {\it et al.},
\textit{Mission design of LiteBIRD},
\href{https://doi.org/10.1007/s10909-013-0996-1}{J.\ Low.\ Temp.\ Phys.\  {\bf 176}, 733 (2014)}
[\href{https://arxiv.org/abs/arXiv:1311.2847}{1311.2847}].

\bibitem{Ade:2018sbj} 
J.~Aguirre {\it et al.} [Simons Observatory Collaboration],
\textit{The Simons Observatory: Science goals and forecasts},
\href{https://doi.org/10.1088/1475-7516/2019/02/056}{JCAP {\bf 1902}, 056 (2019)}
[\href{https://arxiv.org/abs/arXiv:1808.07445}{1808.07445}].

\bibitem{Martin:2013tda} 
J.~Martin, C.~Ringeval and V.~Vennin,
\textit{Encyclop\ae dia Inflationaris},
\href{https://doi.org/10.1016/j.dark.2014.01.003}{Phys.\ Dark Univ.\  {\bf 5-6}, 75 (2014)}
[\href{https://arxiv.org/abs/arXiv:1303.3787}{1303.3787}].

\bibitem{Callan:1970ze} 
C.~G.~Callan, Jr., S.~R.~Coleman and R.~Jackiw,
\textit{A New improved energy - momentum tensor},
\href{https://doi.org/10.1016/0003-4916(70)90394-5}{Annals Phys.\  {\bf 59}, 42 (1970)}.

\bibitem{Freedman:1974ze} 
D.~Z.~Freedman and E.~J.~Weinberg,
\textit{The Energy-Momentum Tensor in Scalar and Gauge Field Theories}, \href{https://doi.org/10.1016/0003-4916(74)90040-2}{Annals Phys.\  {\bf 87}, 354 (1974)}.

\bibitem{Buchbinder:1992rb} 
I.~L.~Buchbinder, S.~D.~Odintsov and I.~L.~Shapiro,
\textit{Effective action in quantum gravity},
Bristol, UK: IOP (1992) 413 p.

\bibitem{Salopek:1988qh} 
D.~S.~Salopek, J.~R.~Bond and J.~M.~Bardeen,
\textit{Designing Density Fluctuation Spectra in Inflation},
\href{https://doi.org/10.1103/PhysRevD.40.1753}{Phys.\ Rev.\ D {\bf 40}, 1753 (1989)}.

\bibitem{Bezrukov:2007ep} 
F.~L.~Bezrukov and M.~Shaposhnikov,
\textit{The Standard Model Higgs boson as the inflaton},
\href{https://doi.org/10.1016/j.physletb.2007.11.072}{Phys.\ Lett.\ B {\bf 659}, 703 (2008)}
[\href{https://arxiv.org/abs/arXiv:0710.3755}{0710.3755}].

\bibitem{Bauer:2008zj} 
F.~Bauer and D.~A.~Demir,
\textit{Inflation with Non-Minimal Coupling: Metric versus Palatini Formulations},
\href{https://doi.org/10.1016/j.physletb.2008.06.014}{Phys.\ Lett.\ B {\bf 665}, 222 (2008)}
[\href{https://arxiv.org/abs/arXiv:0803.2664}{0803.2664}].

\bibitem{York:1972sj} 
J.~W.~York, Jr.,
\textit{Role of conformal three geometry in the dynamics of gravitation},
\href{https://doi.org/10.1103/PhysRevLett.28.1082}{Phys.\ Rev.\ Lett.\  {\bf 28}, 1082 (1972)}.

\bibitem{Tenkanen:2017jih} 
T.~Tenkanen,
\textit{Resurrecting Quadratic Inflation with a non-minimal coupling to gravity},
\href{https://doi.org/10.1088/1475-7516/2017/12/001}{JCAP {\bf 1712}, no. 12, 001 (2017)}
[\href{https://arxiv.org/abs/arXiv:1710.02758}{1710.02758}].

\bibitem{Rasanen:2017ivk} 
S.~Rasanen and P.~Wahlman,
\textit{Higgs inflation with loop corrections in the Palatini formulation},
\href{https://doi.org/10.1088/1475-7516/2017/11/047}{JCAP {\bf 1711}, no. 11, 047 (2017)}
[\href{https://arxiv.org/abs/arXiv:1709.07853}{1709.07853}].

\bibitem{Racioppi:2017spw} 
A.~Racioppi,
\textit{Coleman-Weinberg linear inflation: metric vs. Palatini formulation},
\href{https://doi.org/10.1088/1475-7516/2017/12/041}{JCAP {\bf 1712}, no. 12, 041 (2017)}
[\href{https://arxiv.org/abs/arXiv:1710.04853}{1710.04853}].


\bibitem{Tamanini:2010uq} 
N.~Tamanini and C.~R.~Contaldi,
\textit{Inflationary Perturbations in Palatini Generalised Gravity},
\href{https://doi.org/10.1103/PhysRevD.83.044018}{Phys.\ Rev.\ D {\bf 83}, 044018 (2011)}
[\href{https://arxiv.org/abs/arXiv:1010.0689}{1010.0689}].


\bibitem{Kallosh:2013tua} 
R.~Kallosh, A.~Linde and D.~Roest,
\textit{Universal Attractor for Inflation at Strong Coupling},
\href{https://doi.org/10.1103/PhysRevLett.112.011303}{Phys.\ Rev.\ Lett.\  {\bf 112}, no. 1, 011303 (2014)}
[\href{https://arxiv.org/abs/arXiv:1310.3950}{1310.3950}].

\bibitem{Barrie:2016rnv} 
N.~D.~Barrie, A.~Kobakhidze and S.~Liang,
\textit{Natural Inflation with Hidden Scale Invariance},
\href{https://doi.org/10.1016/j.physletb.2016.03.056}{Phys.\ Lett.\ B {\bf 756}, 390 (2016)}
[\href{https://arxiv.org/abs/arXiv:1602.04901}{1602.04901}].


\bibitem{Kannike:2015kda} 
K.~Kannike, A.~Racioppi and M.~Raidal,
\textit{Linear inflation from quartic potential},
\href{https://doi.org/10.1007/JHEP01(2016)035}{JHEP {\bf 1601}, 035 (2016)}
[\href{https://arxiv.org/abs/arXiv:1509.05423}{1509.05423}].


\bibitem{Artymowski:2016dlz} 
M.~Artymowski and A.~Racioppi,
\textit{Scalar-tensor linear inflation},
\href{https://doi.org/10.1088/1475-7516/2017/04/007}{JCAP {\bf 1704}, no. 04, 007 (2017)}
[\href{https://arxiv.org/abs/arXiv:1610.09120}{1610.09120}].

\bibitem{Jinno:2019und} 
R.~Jinno, M.~Kubota, K.~y.~Oda and S.~C.~Park,
\textit{Higgs inflation in metric and Palatini formalisms: Required suppression of higher dimensional operators},
\href{https://arxiv.org/abs/arXiv:1904.05699}{1904.05699}.

\bibitem{Rubio:2019ypq} 
J.~Rubio and E.~S.~Tomberg,
\textit{Preheating in Palatini Higgs inflation},
\href{https://doi.org/10.1088/1475-7516/2019/04/021}{JCAP {\bf 1904}, no. 04, 021 (2019)}
[\href{https://arxiv.org/abs/arXiv:1902.10148}{1902.10148}].


\bibitem{Enckell:2018kkc} 
V.~M.~Enckell, K.~Enqvist, S.~Rasanen and E.~Tomberg,
\textit{Higgs inflation at the hilltop},
\href{https://doi.org/10.1088/1475-7516/2018/06/005}{JCAP {\bf 1806}, no. 06, 005 (2018)}
[\href{https://arxiv.org/abs/arXiv:1802.09299}{1802.09299}].


\bibitem{Bostan:2018evz} 
N.~Bostan, \"{O}.~G\"{u}lery\"{u}z and V.~N.~\c{S}eno\u{g}uz,
\textit{Inflationary predictions of double-well, Coleman-Weinberg, and hilltop potentials with non-minimal coupling},
\href{https://doi.org/10.1088/1475-7516/2018/05/046}{JCAP {\bf 1805}, no. 05, 046 (2018)}
[\href{https://arxiv.org/abs/arXiv:1802.04160}{1802.04160}].

\bibitem{Zee:1978wi} 
A.~Zee,
\textit{A Broken Symmetric Theory of Gravity},
\href{https://doi.org/10.1103/PhysRevLett.42.417}{Phys.\ Rev.\ Lett.\  {\bf 42}, 417 (1979)}.

\bibitem{Lyth:2009zz} 
D.~H.~Lyth and A.~R.~Liddle,
\textit{The primordial density perturbation: Cosmology, inflation and the origin of structure},
Cambridge, UK: Cambridge Univ. Pr. (2009).

\bibitem{Linde:2011nh} 
A.~Linde, M.~Noorbala and A.~Westphal,
\textit{Observational consequences of chaotic inflation with nonminimal coupling to gravity}, \href{https://doi.org/10.1088/1475-7516/2011/03/013}{JCAP {\bf 1103}, 013 (2011)}
[\href{https://arxiv.org/abs/arXiv:1101.2652}{1101.2652}].


\bibitem{Liddle:2003as} 
A.~R.~Liddle and S.~M.~Leach,
\textit{How long before the end of inflation were observable perturbations produced?},
\href{https://doi.org/10.1103/PhysRevD.68.103503}{Phys.\ Rev.\ D  {\bf 68}, 103503 (2003)}
[\href{https://arxiv.org/abs/arXiv:astro-ph/0305263}{0305263}].

\bibitem{Tenkanen:2016twd} 
T.~Tenkanen,
\textit{Feebly Interacting Dark Matter Particle as the Inflaton}, \href{https://doi.org/10.1007/JHEP09(2016)049}{JHEP {\bf 1609}, 049 (2016)}
[\href{https://arxiv.org/abs/arXiv:1607.01379}{1607.01379}].

\bibitem{Alanne:2016mpa} 
T.~Alanne, F.~Sannino, T.~Tenkanen and K.~Tuominen,
\textit{Inflation and pseudo-Goldstone Higgs boson}, \href{https://doi.org/10.1103/PhysRevD.95.035004}{Phys.\ Rev.\ D {\bf 95}, no. 3, 035004 (2017)}
[\href{https://arxiv.org/abs/arXiv:1611.04932}{1611.04932}].



\bibitem{Goldstone:1961eq} 
J.~Goldstone,
\textit{Field Theories with Superconductor Solutions},
\href{https://doi.org/10.1007/BF02812722}{Nuovo Cim.\  {\bf 19}, 154 (1961)}.


\bibitem{Vilenkin:1994pv} 
A.~Vilenkin,
\textit{Topological inflation},
\href{https://doi.org/10.1103/PhysRevLett.72.3137}{Phys.\ Rev.\ Lett.\  {\bf 72}, 3137 (1994)}
[\href{https://arxiv.org/abs/arXiv:hep-th/9402085}{9402085}].

\bibitem{Linde:1994wt} 
A.~D.~Linde and D.~A.~Linde,
\textit{Topological defects as seeds for eternal inflation},
\href{https://doi.org/10.1103/PhysRevD.50.2456}{Phys.\ Rev.\ D {\bf 50}, 2456 (1994)}
[\href{https://arxiv.org/abs/arXiv:hep-th/9402115}{9402115}].

\bibitem{Destri:2007pv} 
C.~Destri, H.~J.~de Vega and N.~G.~Sanchez,
\textit{MCMC analysis of WMAP3 and SDSS data points to broken symmetry inflaton potentials and provides a lower bound on the tensor to scalar ratio},
\href{https://doi.org/10.1103/PhysRevD.77.043509}{Phys.\ Rev.\ D {\bf 77}, 043509 (2008)}
[\href{https://arxiv.org/abs/arXiv:astro-ph/0703417}{0703417}].


\bibitem{Okada:2014lxa} 
N.~Okada, V.~N.~\c{S}eno\u{g}uz and Q.~Shafi,
\textit{The Observational Status of Simple Inflationary Models: an Update},
\href{https://doi.org/10.3906/fiz-1505-7}{Turk.\ J.\ Phys.\  {\bf 40}, no. 2, 150 (2016)}
[\href{https://arxiv.org/abs/arXiv:1403.6403}{1403.6403}].

\bibitem{Almeida:2018oid} 
J.~P.~B.~Almeida, N.~Bernal, J.~Rubio and T.~Tenkanen,
\textit{Hidden Inflaton Dark Matter},
\href{https://doi.org/10.1088/1475-7516/2019/03/012}{JCAP {\bf 1903}, 012 (2019)}
[\href{https://arxiv.org/abs/arXiv:1811.09640}{1811.09640}].


\bibitem{Tenkanen:2019jiq} 
T.~Tenkanen,
\textit{Minimal Higgs inflation with an $R^2$ term in Palatini gravity},
\href{https://doi.org/10.1103/PhysRevD.99.063528}{Phys.\ Rev.\ D {\bf 99}, no. 6, 063528 (2019)}
[\href{https://arxiv.org/abs/arXiv:1901.01794}{1901.01794}].


\bibitem{Takahashi:2018brt} 
T.~Takahashi and T.~Tenkanen,
\textit{Towards distinguishing variants of non-minimal inflation},
\href{https://doi.org/10.1088/1475-7516/2019/04/035}{JCAP {\bf 1904}, no. 04, 035 (2019)}
[\href{https://arxiv.org/abs/arXiv:1812.08492}{1812.08492}].

\bibitem{Burns:2016ric} 
D.~Burns, S.~Karamitsos and A.~Pilaftsis,
\textit{Frame-Covariant Formulation of Inflation in Scalar-Curvature Theories},
\href{https://doi.org/10.1016/j.nuclphysb.2016.04.036}{Nucl.\ Phys.\ B {\bf 907}, 785 (2016)} [\href{https://arxiv.org/abs/arXiv:1603.03730}{1603.03730}].

\bibitem{Kaiser:1994vs} 
D.~I.~Kaiser,
\textit{Primordial spectral indices from generalized Einstein theories},
\href{https://doi.org/10.1103/PhysRevD.52.4295}{Phys.\ Rev.\ D {\bf 52}, 4295 (1995)} [\href{https://arxiv.org/abs/arXiv:astro-ph/9408044}{9408044}].

\bibitem{Cerioni:2009kn} 
A.~Cerioni, F.~Finelli, A.~Tronconi and G.~Venturi,
\textit{Inflation and Reheating in Induced Gravity},
\href{https://doi.org/10.1016/j.physletb.2009.10.066}{Phys.\ Lett.\ B {\bf 681}, 383 (2009)} [\href{https://arxiv.org/abs/arXiv:0906.1902}{0906.1902}].

\bibitem{Tronconi:2017wps} 
A.~Tronconi,
\textit{Asymptotically Safe Non-Minimal Inflation},
\href{https://doi.org/10.1088/1475-7516/2017/07/015}{JCAP {\bf 1707}, no. 07, 015 (2017)} [\href{https://arxiv.org/abs/arXiv:1704.05312}{1704.05312}].


\bibitem{Izawa:1996dv} 
K.-I.~Izawa and T.~Yanagida,
\textit{Natural new inflation in broken supergravity},
\href{https://doi.org/10.1016/S0370-2693(96)01638-3}{Phys.\ Lett.\ B {\bf 393}, 331 (1997)}.



\bibitem{Kawasaki:2003zv} 
M.~Kawasaki, M.~Yamaguchi and J.~Yokoyama,
\textit{Inflation with a running spectral index in supergravity},
\href{https://doi.org/10.1103/PhysRevD.68.023508}{Phys.\ Rev.\ D {\bf 68}, 023508 (2003)} [\href{https://arxiv.org/abs/arXiv:hep-ph/0304161}{0304161}].

\bibitem{Senoguz:2004ky} 
V.~N.~Senoguz and Q.~Shafi,
\textit{New inflation, preinflation, and leptogenesis},
\href{https://doi.org/10.1016/j.physletb.2004.05.077}{Phys.\ Lett.\ B {\bf 596}, 8 (2004)} [\href{https://arxiv.org/abs/arXiv:hep-ph/0403294}{0403294}].







\end{thebibliography}
\end{document}